\documentclass[journal]{IEEEtran}

\usepackage{amsmath,amsfonts}
\usepackage{algorithmic}
\usepackage{algorithm}
\usepackage{array}
\usepackage{textcomp}
\usepackage{stfloats}
\usepackage{url}
\usepackage{verbatim}
\usepackage{graphicx}
\usepackage{cite}
\usepackage{tablefootnote}
\usepackage{xspace, subfigure, amssymb}
\usepackage[nolist,nohyperlinks]{acronym}
\usepackage{overpic}
\usepackage{lipsum}
\usepackage{xcolor}
\usepackage{multirow}

\newcommand{\chaptername}{Chap.}
\newcommand{\sectionname}{Sect.}
\newcommand{\sectionsname}{Sects.}

\newcommand{\tablesname}{Tables}

\renewcommand{\tablename}{Table}

\newcommand{\meter}{\,\mathrm{m}}
\newcommand{\km}{\,\mathrm{km}}

\newcommand{\cm}{\,\mathrm{cm}}
\newcommand{\dB}{\,\mathrm{dB}}
\newcommand{\dBm}{\,\mathrm{dBm}}

\newcommand{\kHz}{\,\mathrm{kHz}}
\newcommand{\MHz}{\,\mathrm{MHz}}
\newcommand{\GHz}{\,\mathrm{GHz}}
\newcommand{\seconds}{\,\mathrm{s}}
\newcommand{\ms}{\,\mathrm{ms}}

\def\tr{\mathrm{tr}}
\def\Htran{\mathsf{H}}
\def\Ttran{\mathsf{T}}
\def\imagunit{\mathsf{j}} 
\newcommand{\vect}[1]{\boldsymbol{\mathbf{#1}}}

\newcommand{\trace}[1]{\mathrm{tr}\left({#1}\right)}
\newcommand{\diag}[1]{\mathrm{diag}\left({#1}\right)}
\newcommand{\rank}[1]{\mathrm{rank}\left({#1}\right)}
\newcommand{\Ex}[1]{\mathbb{E}\left\{{#1}\right\}}
\def\sinc{\mathrm{sinc}}

\renewcommand{\mod}[1]{\mathrm{mod}\left({#1}\right)}

\begin{document}

\title{\acs{MMSE} Channel Estimation in Large-Scale \acs{MIMO}: Improved Robustness with Reduced Complexity}

\author{Giacomo Bacci, \emph{Senior~Member, IEEE}, Antonio Alberto D'Amico, \emph{Senior~Member, IEEE}, and\\Luca Sanguinetti, \emph{Senior~Member, IEEE}

\thanks{A preliminary version of this article can be found in \cite{Damico_2023}.}
\thanks{Manuscript received 3 April 2024; revised 2 August 2024; accepted 25 September 2024. Date of publication XX MONTH 202Y; date of current version XX MONTH 202Y. The
  associate editor coordinating the review of this article and approving it for publication was W. Cheng. \emph{(Corresponding author: Giacomo Bacci.)}}
\thanks{G.~Bacci, A.A.~D'Amico and L.~Sanguinetti are with the Dipartimento di Ingegneria dell'Informazione, University of Pisa, 56122, Pisa, Italy (email: giacomo.bacci@unipi.it, antonio.damico@unipi.it, luca.sanguinetti@unipi.it). They are partially supported by the Italian Ministry of Education and Research (MUR) in the framework of the FoReLab project (Departments of Excellence).}
\thanks{Color versions of one or more figures in this article are available at https://doi.org/10.1109/TWC.2024.3470124.}
\thanks{Digital Object Identifier 10.1109/TWC.2024.3470124}

}

\maketitle

\begin{abstract}
Large-scale \acs{MIMO} systems with a massive number $N$ of individually controlled antennas pose significant challenges for \ac{MMSE} channel estimation, based on uplink pilots. The major ones arise from the computational complexity, which scales with $N^{3}$, and from the need for accurate knowledge of the channel statistics. This paper aims to address both challenges by introducing reduced-complexity channel estimation methods that achieve the performance of \ac{MMSE} in terms of estimation accuracy and uplink spectral efficiency while demonstrating improved robustness in practical scenarios where channel statistics must be estimated. This is achieved by exploiting the inherent structure of the spatial correlation matrix induced by the array geometry. Specifically, we use a Kronecker decomposition for uniform planar arrays and a well-suited circulant approximation for uniform linear arrays. By doing so, a significantly lower computational complexity is achieved, scaling as $N\sqrt{N}$ and $N \log N$ for squared planar arrays and linear arrays, respectively.
\end{abstract}


\begin{acronym}[DVB-S2X]
\acro{AWGN}{additive white Gaussian noise}
\acro{BS}{base station}
\acro{CDF}{cumulative distribution function}
\acro{CS}{compressed sensing}
\acro{CSI}{channel state information}
\acro{DL}{downlink}
\acro{DFT}{discrete Fourier transform}
\acro{ELAA}{extremely large aperture array}
\acro{EM}{electromagnetic}
\acro{KBA}{Kronecker-based approximation}
\acro{KPI}{key performance indicator}
\acro{hMIMO}{holographic \acs{MIMO}}
\acro{mMIMO}{massive \acs{MIMO}}
\acro{i.i.d.}{independent and identically distributed}
\acro{ISO}{isotropic}
\acro{LoS}{line-of-sight}
\acro{MAI}{multiple access interference}
\acro{MIMO}{multiple-input multiple-output}
\acro{MR}{maximum ratio}
\acro{MSE}{mean square error}
\acro{MMSE}{minimum mean square error}
\acro{NMSE}{normalized mean square error}
\acro{NMSE}{normalized mean square estimation error}
\acro{NSAE}{normalized squared approximation error}
\acro{OMP}{orthogonal matching pursuit}
\acro{LoS}{line-of-sight}
\acro{LS}{least-squares}
\acro{NKP}{nearest Kronecker product}
\acro{NLoS}{non-line-of-sight}
\acro{RF}{radio-frequency}
\acro{RHS}{right-hand side}
\acro{RIS}{reflective intelligent surface}
\acro{RS-LS}{reduced-subspace \acs{LS}}
\acro{RZF}{regularized zero forcing}
\acro{SE}{spectral efficiency}
\acro{SINR}{signal-to-interference-plus-noise ratio}
\acro{SNR}{signal-to-noise ratio}
\acro{SVD}{singular value decomposition}
\acro{UaF}{use-and-then-forget}
\acro{UPA}{uniform planar array}
\acro{UL}{uplink}
\acro{ULA}{uniform linear array}
\acro{UE}{user equipment}
\acro{MIMO}{multiple-input multiple-output}
\acro{UaF}{use-and-then-forget}
\acro{pdf}{probability density function}
\end{acronym}

\acresetall

\section{Introduction and motivation}\label{sec:intro}


Communications theorists are always looking for new technologies to improve the speed and reliability of wireless communications. Among these technologies, the multiple antenna technology has made significant advances, with the latest implementation being massive \ac{MIMO} \cite{massivemimobook}, introduced with the advent of 5G \cite{dsp_nextmMIMO}. Researchers are now exploring improved deployment methods for massive \ac{MIMO}, incorporating more antennas and optimized signal processing algorithms to take advantage of its potential benefits. This evolution of massive \ac{MIMO} has been referred to as massive \ac{MIMO} 2.0 \cite{sanguinettiTCOM2020}, leading to the exploration of \ac{MIMO} systems with extremely larger antenna arrays. This is referred to as X-MIMO in industry terminology. Various terms, such as \acp{ELAA} \cite{dsp_nextmMIMO}, extremely large-scale \ac{MIMO} (XL-MIMO) \cite{Niyato_2023}, and ultra-massive \ac{MIMO} (UM-MIMO) \cite{bjornson20246g}, have also been suggested in academic literature. 
In this paper, we simply use the large-scale \ac{MIMO} terminology.

Accurate \ac{CSI} is crucial for the efficient use of large-scale \ac{MIMO}, enabling precise transmission beamforming in \ac{DL} and coherent signal combining in \ac{UL}. The basic way to acquire \ac{CSI} in time-division duplexing systems is to transmit a predefined pilot sequence and estimate the channel coefficients at the \ac{BS}. However, performing \ac{MMSE} channel estimation in a large-scale \ac{MIMO} system with a massive number $N$ of independent \ac{RF} chains (e.g., in the order of thousands) poses critical challenges in the implementation of the \ac{MMSE} estimator (whose computational complexity may scale as $N^3$ due to matrix inversion, e.g., \cite{AnTutorialPartI_2023, an2022}) and in the acquisition of the channel statistics, i.e., the spatial correlation matrices. In practical communication scenarios, the latter are not known a priori and must be estimated based on observed channel measurements \cite{massivemimobook}. This is crucial for designing robust and efficient large-scale \ac{MIMO} systems \cite{sanguinettiTCOM2020}. An alternative is to use the \ac{LS} estimator \cite{Liu2014}, which requires neither matrix inversion nor statistical information. However, its performance is significantly inferior to \ac{MMSE}, e.g., \cite{massivemimobook}. The objective of this paper is to tackle this challenge by developing channel estimation schemes that achieve accuracy levels comparable to the \ac{MMSE} estimator, while also offering reduced complexity and enhanced robustness to the imperfect knowledge of channel statistics. 



\subsection{Related work}\label{intro:related}

A variety of schemes exists in the literature to achieve the same performance of \ac{MMSE} while handling complexity that scales with the number of antennas. These include, for example, methods based on exhaustive search \cite{Dai2006}, hierarchical search \cite{Xiao2016, Zhang2017, Noh2017}, \ac{CS} \cite{Wan2021, Cui2022, Ghermezcheshmeh2023}, and tensor decomposition \cite{Qin18, Cheng19, Zhang21}. The exhaustive method in \cite{Dai2006} exploits a sequence of training symbols, whose overhead becomes prohibitively high as the number $N$ of antennas increases. To overcome this issue, a hierarchical search based on a predefined codebook is proposed in \cite{Xiao2016, Zhang2017, Noh2017}. However, while significantly reducing the complexity, the limited codebook size severely affects the estimation accuracy. Alternative methods to reduce the pilot overhead take advantage of \ac{CS} techniques, which exploit the fact that the propagation channels of most practical communication scenarios can be sparsely represented in the angular domain. 
A parametric approach is used in \cite{Ghermezcheshmeh2023} for \ac{LoS} communications. A different approach is pursued in \cite{demir2022WCL}, which exploits the high rank deficiency observed in the spatial correlation matrices caused by the large-scale \ac{MIMO} geometry \cite{pizzo2022} to derive a subspace-based channel estimation approach. The complexity saving is due to a compact eigenvalue decomposition, which allows the proposed \ac{RS-LS} estimator to attain the \ac{MMSE} performance when the pilot \ac{SNR} becomes significantly large. However, this might not hold true, especially for practical transmitter-receiver distances. 

The vast majority of the aforementioned literature assumes that the channel statistics are perfectly known. However, this assumption is questionable because the matrix dimensions grow with $N$ and the statistics change over time (e.g., due to mobility). Practical covariance estimates are imperfect because the number of observations may be comparable to $N$. One promising approach to estimating a large-dimensional covariance matrix with a small number of observations is to regularize the sample covariance matrix \cite{Bjornson2016,massivemimobook}. This makes it scenario-dependent.

\subsection{Main contributions}\label{intro:contributions}

This paper focuses on the development of channel estimation schemes that, compared to the optimal MMSE estimator, offer both lower computational complexity and improved robustness in the presence of imperfect knowledge of channel statistics. This is accomplished by exploiting the inherent structure of the spatial correlation matrix shaped by the array geometry, which is known in a given deployment scenario. For a \ac{UPA}, we use the Kronecker product decomposition \cite{VanLoan1992, VanLoan2000} to effectively separate the horizontal and vertical components within the array. For a \ac{ULA}, we use an appropriate circulant approximation \cite{Pearl1971,Pearl1973,Wakin2017}, which allows the application of the discrete Fourier transform (DFT). 
Numerical results are used to evaluate the effectiveness of the proposed methods in terms of \ac{NMSE} and \ac{SE} in the uplink with different combining schemes. The results show that the proposed schemes achieve performance levels comparable to those with the \ac{MMSE} estimator, while significantly reducing computational complexity. Specifically, the overall computational load, including the estimation of the spatial correlation matrices, scales as $N\sqrt{N}$ and $N \log N$ for squared planar arrays and linear arrays, respectively. Moreover, the schemes exhibit improved robustness in scenarios with imperfect knowledge of channel statistics, making them more suitable for dynamic environments. 
Remarkably, this is achieved without any regularization factor, making the schemes applicable without the need for fine-tuning according to the specific scenarios.

\subsection{Paper outline and notation}\label{intro:structure}
The remainder of the paper is structured as follows. \sectionname~\ref{sec:upa_system_model} presents the system and channel models, whereas \sectionname~\ref{sec:perfect_knowledge} revises the channel estimation problem in the presence of perfect knowledge of the channel statistics. \sectionsname~\ref{sec:perfect_upa} and \ref{sec:perfect_ula} derive low-complexity schemes for \ac{UPA} and \ac{ULA} scenarios, respectively. \sectionname~\ref{sec:imperfect_knowledge} addresses the problem of estimating the channel statistics. \sectionsname~\ref{sec:complexity} and \ref{sec:ul_sp} investigate the complexity and the performance of the proposed schemes in terms of achievable uplink \ac{SE}. \sectionname~\ref{sec:conclusion} concludes the paper.

Matrices and vectors are denoted by boldface uppercase and lowercase letters, respectively.
The notation $[\vect{A}]_{i,k}$ is used to indicate the $(i,k)$th entry of the enclosed matrix $\vect{A}$,
and $\vect{A}=\diag{a(n); n=1,\dots,N}$ denotes an $N \times N$ diagonal matrix with entries $a(n)$ along its main diagonal.
$\|\vect{x}\|$ denotes the Euclidean norm of vector $\vect{x}$, and
$\|\vect{A}\|_\mathsf{F}$ denotes the Frobenius norm of matrix $\vect{A}$.
Trace and rank of a matrix $\vect{A}$ are denoted by $\trace{\vect{A}}$ and $\rank{\vect{A}}$, respectively,
whereas $\vect{A}^\Ttran$ and $\vect{A}^\Htran$ are the transpose and the conjugate
transpose of $\vect{A}$, respectively, and $\imagunit$ denotes the imaginary unit. The $N \times N$ identity
matrix and the all-zero vector with $N$ elements are denoted by $\vect{I}_N$ and $\vect{0}_N$, respectively. We use $\delta(\cdot)$ to denote the Dirac delta function, $\mod{\cdot,\cdot}$ to denote the modulus operation, $\lfloor\cdot\rfloor$ to truncate the argument,  $\otimes$ to indicate the Kronecker product, and $\Re(\cdot)$ to represent the real part of a complex number.

\subsection{Reproducible research}\label{intro:code}

The \textsc{MatLab} code used to obtain the simulation results will be made available to the interested readers upon request.

\section{System and channel model}\label{sec:upa_system_model}

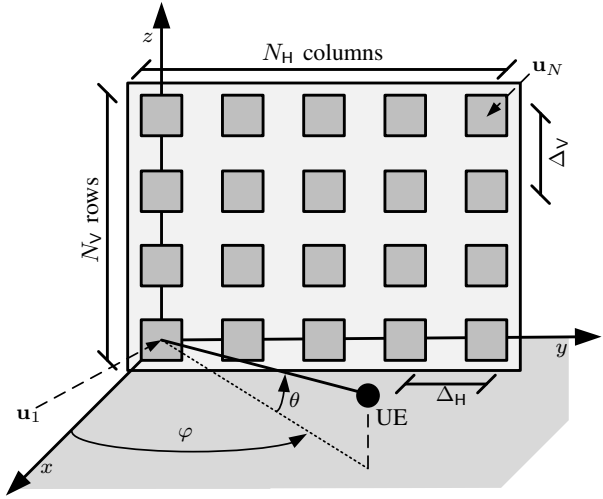
\begin{figure}[t!]
  \begin{center}
    \begin{overpic}[width=0.9\columnwidth]{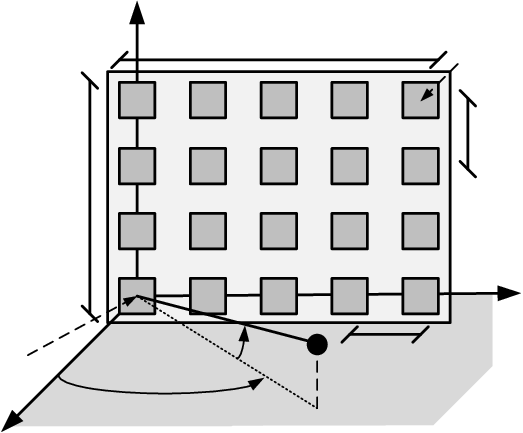}
      \put(6,4){\footnotesize{$x$}}
      \put(92,24){\footnotesize{$y$}}
      \put(23,76){\footnotesize{$z$}}
      \put(43,73){\small{$N_\mathsf{H}$ columns}}
      \put(72,16){\footnotesize{$\Delta_\mathsf{H}$}}
      \put(13,39){\rotatebox{+90}{\small{$N_\mathsf{V}$ rows}}}
      \put(91,55){\rotatebox{+90}{\footnotesize{$\Delta_\mathsf{V}$}}}
      \put(62,12){\small{\acs{UE}}}
      \put(29,10){\footnotesize{$\varphi$}}
      \put(47.5,15){\footnotesize{$\theta$}}
      \put(2,13){\footnotesize{$\vect{u}_1$}}
      \put(88,71){\footnotesize{$\vect{u}_N$}}
    \end{overpic}
\caption{Diagram of the \acs{UPA} located in the $yz$-plane, with a planar wave impinging with elevation $\theta$ and azimuth $\varphi$.}
\label{fig01}
\end{center}
\end{figure}

We consider a large-scale \ac{MIMO} system with $K$ active \acp{UE}. The \ac{BS} is equipped with a \ac{UPA} located in the $yz$ plane, and consisting of $N$ independent \ac{RF} chains, arranged into $N_\mathsf{V}$ rows, each hosting $N_\mathsf{H}$ antennas: $N=N_\mathsf{H}N_\mathsf{V}$.\footnote{The analysis is valid for any orientation of the \ac{UPA} with respect to the reference system, and includes both horizontal and vertical \acp{ULA},
by setting $N_\mathsf{V}=1$ and $N_\mathsf{H}=1$, respectively.} As illustrated in \figurename~\ref{fig01}, the inter-element spacing across 
horizontal and vertical directions is $\Delta_\mathsf{H}$ and $\Delta_\mathsf{V}$, respectively \cite[\figurename~1]{Sanguinetti2020}. 
The location of the $n$th antenna, with $1 \le n \le N$, with respect to the origin is $\vect{u}_n = [ 0, i(n) \Delta_\mathsf{H}, j(n) \Delta_\mathsf{V} ]^{\Ttran}$, 
where $i(n)=\mod{n-1, N_\mathsf{H}}$ and $j(n)=\left\lfloor\left(n-1\right)/N_\mathsf{H}\right\rfloor$
are the horizontal and vertical indices of element $n$.

We call $\vect{h}_k\in \mathbb{C}^N$ the channel vector between the single-antenna \ac{UE} $k$ and the \ac{BS}, and model it as~\cite{Sayeed2002a}\begin{align} \label{eq:channel1}
  \vect{h}_k =  \iint_{-\pi/2}^{\pi/2} g_k(\varphi,\theta) \vect{a}(\varphi,\theta) d\varphi d\theta  
\end{align}
where $\vect{a}(\varphi,\theta)$ is
the array response vector \cite[\sectionname~7.3]{massivemimobook}
\begin{align}\label{eq:array-response}
  \vect{a}(\varphi,\theta) = \left[e^{\imagunit\vect{k}(\varphi,\theta)^{\Ttran}\vect{u}_1},\dots,e^{\imagunit\vect{k}(\varphi,\theta)^{\Ttran}\vect{u}_N}\right]^{\Ttran}
\end{align}
of the \ac{UPA} from azimuth angle $\varphi$ and elevation angle $\theta$, and $ \vect{k}(\varphi, \theta) = \frac{2\pi}{\lambda}\left[\cos(\theta) \cos(\varphi), \, \cos(\theta) \sin(\varphi), \, \sin(\theta)\right]^{\Ttran}$
is the wave vector at wavelength $\lambda$. Also, $g_k(\varphi,\theta)$ is the \emph{angular spreading function} \cite[\sectionname~2.6]{massivemimobook}, which accounts for the local scattering model \cite[\chaptername~7]{molisch2022}.

We consider the conventional block-fading model, where the channel $\vect{h}_k$ is constant within
one time-frequency block and takes independent realizations across blocks from a stationary stochastic distribution.
In accordance with \cite{Sayeed2002a}, we model $g_k(\varphi,\theta)$ as a spatially uncorrelated circularly symmetric Gaussian
stochastic process with cross-correlation
\begin{align} \label{eq:scattering-correlation-model}
  \Ex{g_k(\varphi,\theta) g_k^*(\varphi^\prime,\theta^\prime)} = \beta_k f_k(\varphi,\theta) \delta(\varphi-\varphi^\prime)  \delta(\theta-\theta^\prime)
\end{align}
where $\beta_k$ is the average channel gain and $f_k(\varphi,\theta)$ is the normalized \emph{spatial scattering function} \cite{Sayeed2002a} such that $\iint f_k(\varphi,\theta) d\theta d\varphi  = 1$. By using \eqref{eq:array-response}, the
elements of $\vect{R}_k=\mathbb{E} \{\vect{h}_k \vect{h}_k^{\Htran}\}$ are computed as \cite[\sectionname~7.3.2]{massivemimobook}
\begin{align}\label{eq:3D_local_correlation}
  [\vect{R}_k]_{m,l}  =  \beta_k \iint_{-\pi/2}^{\pi/2} e^{\imagunit\vect{k}(\varphi,\theta)^{\Ttran} \left(\vect{u}_m - \vect{u}_l \right)} f_k(\varphi,\theta)d\varphi d\theta.
\end{align}
In particular, the elements of $\vect{R}_k$ are affected by the 3D local scattering model, that determines the typical values for azimuth and elevation angles, and their respective angular spreads $\sigma_{\varphi}$ and $\sigma_{\theta}$. Measurements indicate that the range of interest for both azimuth and elevation angular spreads in typical cellular scenarios is given by $\left[5^\circ, 15^\circ\right]$, e.g.,\cite{Zhang2017b}.

\section{Channel estimation}\label{sec:perfect_knowledge}

We assume that channel estimation is performed by using orthogonal pilot sequences of length $\tau_p \ge K$. We call 
$\vect{\phi}_{k} \in \mathbb{C}^{\tau_p}$ the pilot sequence used by \ac{UE} $k$ and assume 
$|[\vect{\phi}_{k}]_i|^2 = 1$, so that $\vect{\phi}_{k}^\Htran \vect{\phi}_{k}  = \tau_{p}$. In the absence of pilot contamination, the observation vector is \cite[\sectionname~3]{massivemimobook}
\begin{align}
\vect{y}_k=\tau_p \sqrt{\rho}\vect{h}_{k} + \vect{w}
\end{align}
where $\rho$ is the transmit power and $\vect{w} \sim \mathcal{CN}(\vect{0}_N, \tau_p\sigma^2 {\vect{I}}_N)$.

An arbitrary \emph{linear} estimator of $\mathbf{h}_k$ based on $\vect{y}_k$ takes the form \cite[\sectionname~3]{massivemimobook}
\begin{align}\label{eq:A-generic}
  \widehat{\vect{h}}_k = \vect{A}_k\vect{y}_k
\end{align}
for some deterministic matrix $\mathbf{A}_k\in\mathbb{C}^{N \times N}$ that specifies the estimation scheme.
The \ac{NMSE} of the estimation error $\widetilde{\vect{h}}_k\triangleq\vect{h}_k-\widehat{\vect{h}}_k$ 
can be computed as
\begin{align}\label{eq:mse}
\!\!\mathsf{NMSE} \!=\! 1 - \frac{2 \sqrt{\rho} \tau_p \Re\left( \tr( \vect{R}_{k} \vect{A}_k )  \right) + \rho\tau_p^2 \tr \left( \vect{A}_k \vect{Q}_{k} \vect{A}_k^{\Htran} \right)}{\tr(\vect{R}_{k})}\!\!
\end{align}
where
\begin{align}\label{eq:Q}
  \vect{Q}_k=\frac{1}{\rho\tau_p^2}\Ex{\vect{y}_k \vect{y}_k^\Htran}
  =\vect{R}_k+\frac{1}{\gamma}\vect{I}_N
\end{align}
with $\gamma = \tau_p\rho/\sigma^2$ being the transmit \ac{SNR} (i.e., measured at the UE side), during the channel estimation phase.

\subsection{\acs{MMSE} and \acs{LS} estimators}

The scheme that minimizes \eqref{eq:mse} is the \ac{MMSE} estimator, according to which
\begin{align}\label{eq:A-MMSE}
\vect{A}_k^{\mathsf{MMSE}}= \frac{1}{\tau_p \sqrt{\rho}}{\vect{R}}_k \vect{Q}_{k}^{-1}. 
\end{align}
The \ac{MMSE} estimator is optimal but requires the following operations:
\begin{enumerate}
\item Estimation of the spatial covariance matrix ${\vect{R}}_k$;
\item Computation of the inverse of $\vect{Q}_{k}$;
\item Computation of the channel estimate \begin{align}\widehat{\vect{h}}_k^{\mathsf{MMSE}}=\vect{A}_k^{\mathsf{MMSE}}\vect{y}_k.
\end{align}
\end{enumerate}
The main computational cost comes from the inversion of $\vect{Q}_{k}$, which scales as $\mathcal{O}(N^3)$ if structural properties of the matrix are not exploited. Therefore, an intense computational effort is required when $N$ is large, as envisioned in large-scale \ac{MIMO} \cite{bjornson20246g}.

An alternative is the \ac{LS} estimator with
\begin{align}\label{eq:A-LS}
  \vect{A}_k^{\mathsf{LS}}= \frac{1}{\tau_p \sqrt{\rho}} \vect{I}_N 
\end{align}
which uses no prior information on the channel statistics and array geometry. Unlike the \ac{MMSE} estimator, its computational complexity is $\mathcal{O}(N)$, due to the multiplication between the diagonal matrix $\vect{A}_k^{\mathsf{LS}}$ and $\vect{y}_k$. However, this efficiency comes at the cost of reduced accuracy.

\subsection{\acs{MMSE} estimators for low and rich scattering}

Two estimators for low and rich scattering scenarios are described below. Both take advantage of the array geometry, but apply to two different channel propagation conditions. The first one assumes that propagation 
takes place in a \ac{LoS} scenario with a single plane-wave arriving from $\theta_k$ and $\varphi_k$. Under this hypothesis, 
$\vect{h}_k = g_k(\varphi_k,\theta_k) \vect{a}(\varphi_k,\theta_k) $ and 
${\vect{R}}^{\mathsf{LoS}}_k=\beta_k\vect{a}(\varphi_k,\theta_k)\vect{a}(\varphi_k,\theta_k)^{\Htran}$.
Replacing $\vect{R}_k$ with ${\vect{R}}^{\mathsf{LoS}}_k$ into \eqref{eq:A-MMSE} yields 
$\widehat{\vect{h}}_k^{\mathsf{LoS}} = \vect{A}_k^{\mathsf{LoS}}\vect{y}_k$, where
\begin{align}\label{eq:mmse_estimator_LoS}
  \vect{A}_k^{\mathsf{LoS}} = \frac{1}{\tau_p \sqrt{\rho}}\frac{\beta_k \gamma}{1 +N\beta_k \gamma}
  \vect{a}(\varphi_k,\theta_k)\vect{a}(\varphi_k,\theta_k)^{\Htran}
\end{align}
whose complexity is $\mathcal{C}_{\mathsf{LoS}}=\mathcal{O}(N)$, due to the evaluation of the product between 
$\vect{A}_k^{\mathsf{LoS}}$ in \eqref{eq:mmse_estimator_LoS} and ${\vect y}_k$ (no pre-computation phase is required). 
However, the \ac{LoS}-based estimator works well only when the channel vector is generated by a single plane-wave arriving from
$(\varphi_k,\theta_k)$, whose knowledge must be perfect at the \ac{BS}. 

Conversely, when the propagation scenario is highly scattered, and planar waves arrive uniformly within the angular domain in front of the \ac{UPA}, we can make use of the \ac{ISO} approximation proposed in \cite{demir2022WCL}. According to \cite{demir2022WCL}, ${\vect{R}}^{\mathsf{ISO}}_k= \overline{\vect{U}}\, \overline{\vect{\Lambda}} \,\overline{\vect{U}}^{\Htran}$, where $\overline{\vect{U}}$ is the (reduced-order) eigenvector matrix corresponding to the non-zero eigenvalues in $\overline{\vect{\Lambda}}$, obtained through the \emph{compact} eigenvalue decomposition of ${\vect{R}}^{\mathsf{ISO}}_k$ whose $(m,l)$th entry is \begin{align}
[{\vect{R}}^{\mathsf{ISO}}_k]_{m,l}=\sinc\left(2\sqrt{\delta^2_{\mathsf{H},ml}+\delta^2_{\mathsf{V},ml}}\right)    
\end{align}
with 
\begin{align}
\delta_{\mathsf{H},ml}&=\left[i(m)-i(l)\right]\Delta_\mathsf{H}/\lambda \\ \delta_{\mathsf{V},ml}&=\left[j(m)-j(l)\right]\Delta_\mathsf{V}/\lambda
\end{align}
and $\sinc(x)=\sin(\pi x)/(\pi x)$ \cite{Sanguinetti2020}.
Approximating $\vect{R}_k^{\mathsf{ISO}}$ with $\widetilde{\vect{R}}_k^{\mathsf{ISO}}=\overline{\vect{U}}\overline{\vect{U}}^\Htran$, and
replacing $\vect{R}_k$ with $\widetilde{\vect{R}}_k^{\mathsf{ISO}}$ into \eqref{eq:A-MMSE} yields $\widehat{\vect{h}}_k^{\mathsf{ISO}} = \vect{A}_k^{\mathsf{ISO}}\vect{y}_k$ \cite[Eq. (18)]{demir2022WCL}, with
\begin{align}\label{eq:A-ISO}
  \vect{A}_k^{\mathsf{ISO}} = \frac{1}{\tau_p \sqrt{\rho}} \overline{\vect{U}} \, \overline{\vect{U}}^{\Htran}.
\end{align}
The main advantage of \eqref{eq:A-ISO} is that no matrix estimation and inversion is required, since $\overline{\vect{U}}$ is known and does not depend on \ac{UE} $k$. Therefore, it can be used to precompute and store $\overline{\vect{U}} \, \overline{\vect{U}}^{\Htran}$. Accordingly, the complexity of $\widehat{\vect{h}}_k^{\mathsf{ISO}}$ is only due to the matrix-vector product computation between $\vect{A}_k^{\mathsf{ISO}}$ and $\vect{y}_k$, and is $\mathcal{C}_{\mathsf{ISO}}=\mathcal{O}(N^2)$. 

\section{Reduced-complexity method for \acs{UPA}}\label{sec:perfect_upa}

We first consider a \ac{UPA} and develop a channel estimator that exploits the array geometry and makes use of a Kronecker product approximation of $\vect{R}_k$, in the form
\begin{equation}\label{eq:kron_approx}
\vect{R}_k \approx \vect{K}_1 \otimes \vect{K}_2
\end{equation}
where $\vect{K}_1 \in \mathbb{C}^{N_{\mathsf{V}} \times N_{\mathsf V}}$ and $\vect{K}_2 \in \mathbb{C}^{N_{\mathsf{H}} \times N_{\mathsf H}}$ are suitable matrices. 
The main advantage of using \eqref{eq:kron_approx} is the computational savings in calculating ${\vect{R}}_k \vect{Q}_{k}^{-1}$. In fact, replacing $\vect{R}_k$ with $\vect{K}_1 \otimes \vect{K}_2$ leads to the following approximation of $\vect{A}_k^\mathsf{MMSE}$ in \eqref{eq:A-MMSE}:
\begin{align}\label{eq:A-Kron}
  \vect{A}_k^\mathsf{MMSE} &\approx \vect{A}_k^\mathsf{Kron} = \frac{1}{\tau_p \sqrt{\rho}} 
  \left(\vect{U}_{1}\otimes\vect{U}_{2}\right) 
  \left(\vect{\Lambda}_{1}\otimes\vect{\Lambda}_{2}\right)\nonumber\\
  &\times\left(\vect{\Lambda}_{1}\otimes\vect{\Lambda}_{2} + \frac{1}{\gamma}\vect{I}_N\right)^{-1}
  \left(\vect{U}^{\Htran}_{1}\otimes\vect{U}^{\Htran}_{2}\right)
\end{align}
where $\vect{U}_{1}$ ($\vect{U}_{2}$, respectively) and $\vect{\Lambda}_{1}$ ($\vect{\Lambda}_{2}$, respectively) are the eigenvector and eigenvalue matrices obtained by the spectral decomposition of $\vect{K}_{1}$ ($\vect{K}_{2}$, respectively), i.e., ${\vect{K}}_{1}=\vect{U}_{1} \vect{\Lambda}_{1}\vect{U}_{1}^{\Htran}$ and ${\vect{K}}_{2}=\vect{U}_{2} \vect{\Lambda}_{2}\vect{U}_{2}^{\Htran}$. Compared to the spectral decomposition of $\vect{R}_k$ or the inversion of $\vect{Q}_k$, the spectral decomposition of $\vect{K}_{1}$ and $\vect{K}_{2}$ requires much fewer operations, as will be discussed in \sectionname~\ref{sec:complexity}.

\subsection{Kronecker-based approximation}\label{perfect_upa:kba}
\begin{figure}[t]
  \begin{center}
    \subfigure[$\mathsf{NSAE}_\mathbf{R}$ in \eqref{eq:nsae}.]{
    {\includegraphics[width=\columnwidth]{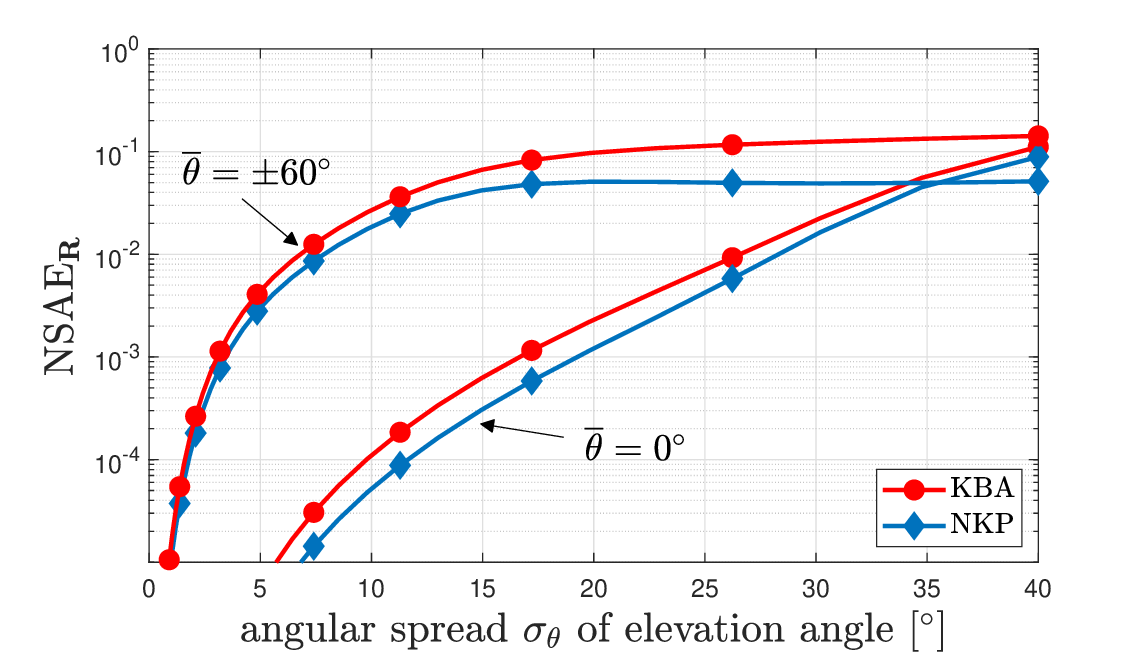}}
    \label{fig02a}}
    \\
    \subfigure[$\mathsf{NSAE}_\mathbf{A}$ in \eqref{eq:nsae_a}.]{
    {\includegraphics[width=\columnwidth]{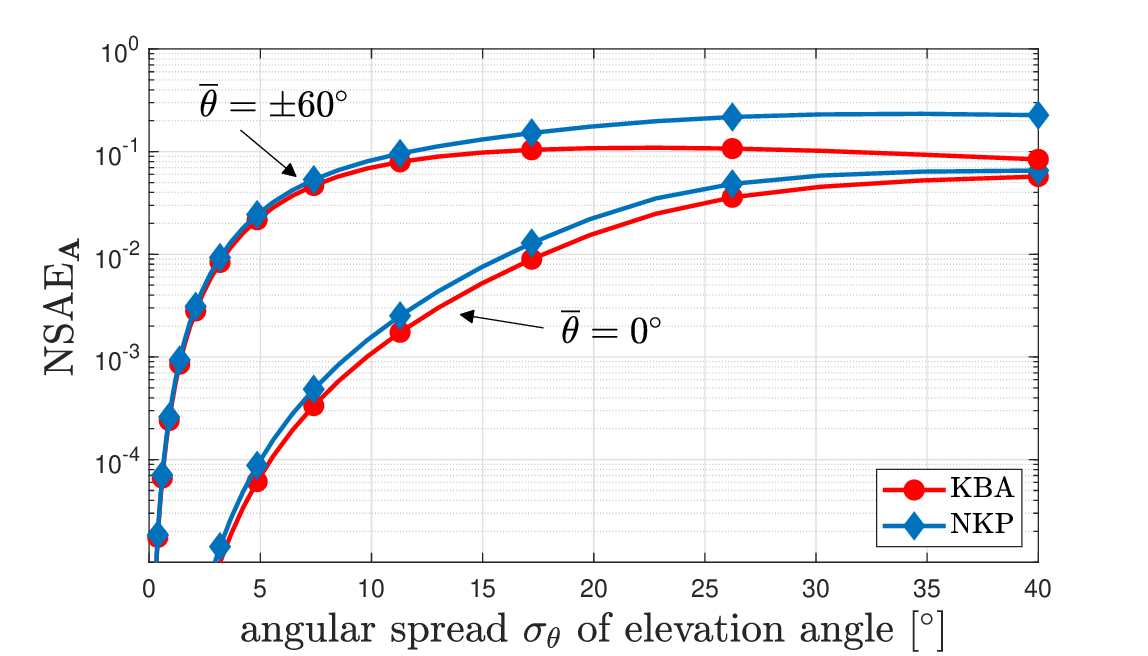}} 
    \label{fig02b}}
    \caption{\acs{NSAE} of the correlation matrix as a function of the elevation angular spread for \acs{KBA} and \ac{NKP}. We consider a \ac{UPA} with $N_\mathsf{H}=N_\mathsf{V}=16$ elements, using $\Delta_\mathsf{H}=\Delta_\mathsf{V}=\lambda/2$ at $3\GHz$. The azimuth angle is $\overline{\varphi}=0^\circ$ with angular spread $\sigma_\varphi=10^\circ$. Both azimuth and elevation angles follow a Gaussian distribution.}
    \label{fig02}
  \end{center}
\end{figure}

We begin by 
rewriting \eqref{eq:3D_local_correlation} as
\begin{align}\label{eq:3D_local_correlation_simplified_2}
  \vect{R}_k  &=  \beta_k \iint_{-\pi/2}^{\pi/2} f_k(\varphi,\theta) 
  \vect{B}_\mathsf{V}(\theta) \otimes \vect{B}_\mathsf{H}(\varphi,\theta) d\varphi d\theta,
\end{align}
where $\vect{B}_\mathsf{H}(\varphi,\theta)\in\mathbb{C}^{N_\mathsf{H}\times N_\mathsf{H}}$ and
$\vect{B}_\mathsf{V}(\theta)\in\mathbb{C}^{N_\mathsf{V}\times N_\mathsf{V}}$ are Hermitian Toeplitz matrices given by
\begin{align}
  \label{eq:kron_aux_H}
  [\vect{B}_\mathsf{H}(\varphi,\theta)]_{m,l}  &=  e^{\imagunit 2\pi (\Delta_\mathsf{H}/\lambda) [i(m)-i(l)]\sin\varphi\cos\theta},\\
  \label{eq:kron_aux_V}
  [\vect{B}_\mathsf{V}(\theta)]_{m,l}          &=  e^{\imagunit 2\pi (\Delta_\mathsf{V}/\lambda) [j(m)-j(l)]\sin\theta}.
\end{align}
Computing the integral \eqref{eq:3D_local_correlation_simplified_2} with respect to $\varphi$ yields
\begin{align}\label{eq:3D_local_correlation_simplified_3}
  \vect{R}_k = \beta_k \int_{-\pi/2}^{\pi/2} \vect{B}_\mathsf{V}(\theta) \otimes \widetilde{\vect{R}}_{\mathsf{H},k}(\theta) d\theta
\end{align}
where
\begin{align}\label{eq:R_H}
  \widetilde{\vect{R}}_{\mathsf{H},k}(\theta) = \int_{-\pi/2}^{\pi/2} f_k(\varphi,\theta) \vect{B}_\mathsf{H}(\varphi,\theta) d\varphi.
\end{align}
In general, the integral expression of $\vect{R}_k$ given in \eqref{eq:3D_local_correlation_simplified_3} cannot be simplified as 
a Kronecker product of two matrices. However, in the specific case where all the plane waves are impinging from the same elevation angle $\overline{\theta}$, i.e., when $f_k(\varphi,\theta)=\tilde{f}_k(\varphi) \delta(\theta-\overline{\theta})$ (which means no angular spread across the elevation angle), we have
\begin{align}\label{eq:KBA}
  \vect{R}_k = \widetilde{\vect{R}}_{\mathsf{V},k}(\overline{\theta}) \otimes \widetilde{\vect{R}}_{\mathsf{H},k}(\overline{\theta})
\end{align}
where
\begin{align}
  \label{eq:R_H_2}
  \widetilde{\vect{R}}_{\mathsf{H},k}(\overline{\theta}) &= \int_{-\pi/2}^{\pi/2} \tilde{f}_k(\varphi) \vect{B}_\mathsf{H}(\varphi,\overline{\theta}) d\varphi\\
  \label{eq:R_V}
  \widetilde{\vect{R}}_{\mathsf{V},k}(\overline{\theta}) &= \beta_k \vect{B}_\mathsf{V}(\overline{\theta}).
\end{align}
Note that the representation of $\vect{R}_k$ in \eqref{eq:KBA} is not unique. Indeed, taking ${\vect{R}}_{\mathsf{H},k} = \alpha \widetilde{\vect{R}}_{\mathsf{H},k}(\overline{\theta})$ and ${\vect{R}}_{\mathsf{V},k} = \alpha^{-1} \widetilde{\vect{R}}_{\mathsf{V},k}(\overline{\theta})$, we have $\vect{R}_k = \vect{R}_{\mathsf{H},k} \otimes \vect{R}_{\mathsf{V},k}$ for any $\alpha \in \mathbb{C}$. We select
\begin{align}
  \label{eq:R_H_Matlab}
  \vect{R}_{\mathsf{H},k} &= [\vect{R}_k]_{1:1:N_\mathsf{H},1:1:N_\mathsf{H}}\\
  \label{eq:R_V_Matlab}
  \vect{R}_{\mathsf{V},k} &=[\vect{R}_k]_{1:N_\mathsf{H}:N,1:N_\mathsf{H}:N}/[\vect{R}_k]_{1,1}
\end{align}
where we have adopted the well-known ``colon notation'' used by \textsc{MatLab}. In the remainder of this paper, the Kronecker decomposition of $\vect{R}_k$ by $\vect{R}_{\mathsf{H},k}$ and $\vect{R}_{\mathsf{V},k}$ will be referred to as the \ac{KBA} of the covariance matrix. The Kronecker product of \eqref{eq:R_H_Matlab} and \eqref{eq:R_V_Matlab} provides an \emph{exact} representation of $\vect{R}_k$ only when $\sigma_\theta=0^\circ$. To assess its robustness for $\sigma_\theta >0^\circ$, \figurename~\ref{fig02a} shows the \ac{NSAE}, defined as
\begin{align}\label{eq:nsae}
  \mathsf{NSAE}_\mathbf{R}= \frac{\left\|\vect{R}_k-\vect{R}_{\mathsf{V},k}\otimes\vect{R}_{\mathsf{H},k}\right\|^2_\mathsf{F}}
  {\|\vect{R}_k\|^2_\mathsf{F}}
\end{align}
as a function of $\sigma_\theta$. We consider a \ac{UPA} with $N_\mathsf{H}=N_\mathsf{V}=16$ elements, using $\Delta_\mathsf{H}=\Delta_\mathsf{V}=\lambda/2$ at $3\GHz$\footnote{Note that, at this carrier frequency, the propagation occurs in the far-field regime \cite{Bacci2023}.} and assume an azimuth angle $\overline{\varphi}=0^\circ$ with angular spread $\sigma_\varphi=10^\circ$. 
Both azimuth and elevation angles follow a Gaussian distribution. For simplicity, we assume $\beta_k=1$.
Two different values of $\overline{\theta}$ are considered: $\overline{\theta}=0^\circ$ 
and $\overline{\theta}=\pm60^\circ$.  As expected, the approximation worsens 
as $\sigma_\theta$ increases. Furthermore, the \ac{NSAE} increases as $\left|\overline{\theta}\right|$ increases, owing to the reduced \ac{UPA} directivity (and hence larger fluctuations in the \ac{KBA}), and, albeit not reported here for the sake of brevity, it also increases as the \ac{UPA} size increases (due to a larger size of the covariance matrix $\vect{R}_k$, which impacts the accuracy of the approximation).

In \figurename~\ref{fig02a}, comparisons are made with the \emph{\acl{NKP}} (\acs{NKP}), derived in \cite{VanLoan1992} by minimizing the \ac{NSAE}. According to \ac{NKP}, the matrix $\vect{R}_k$ is approximated by $ \vect{R}^\star_{\mathsf{V},k} \otimes \vect{R}^\star_{\mathsf{H},k}$, where $ \vect{R}^\star_{\mathsf{H},k}$ and $ \vect{R}^\star_{\mathsf{V},k}$ are obtained as
\begin{equation}
\label{eq:NKP_problem}
\left(\vect{R}^\star_{\mathsf{H},k},\vect{R}^\star_{\mathsf{V},k}\right) = \arg \underset {\left(\vect{X},\vect{Y}\right)} {\min} \left\|\vect{R}_k-\vect{Y} \otimes \vect{X} \right\|^2_\mathsf{F}.
\end{equation}
The \ac{NKP} solution can be found following the steps detailed in \cite{VanLoan2000}. From \figurename~\ref{fig02a}, we see that the difference between \ac{NKP} and \ac{KBA} is relatively small for all the considered values of $\sigma_\theta$. Notice that, despite \eqref{eq:NKP_problem} is the solution that best approximates $\vect{R}_k$, \ac{KBA} performs better than \ac{NKP} as far as the approximation of $\vect{A}^\mathsf{MMSE}_k$ in \eqref{eq:A-MMSE} is concerned. This can be appreciated with the help of \figurename~\ref{fig02b}, which reports 
\begin{align}\label{eq:nsae_a}
  \mathsf{NSAE}_\mathbf{A}= \frac{\left\|\vect{A}^\mathsf{MMSE}_k-\vect{A}_k\right\|^2_\mathsf{F}}
  {\|\vect{A}^\mathsf{MMSE}_k\|^2_\mathsf{F}}
\end{align}
as a function of $\sigma_\theta$ for $\gamma=10\dB$. The matrix $\vect{A}_k$ in \eqref{eq:nsae_a} is obtained from \eqref{eq:A-MMSE} by replacing $\vect{R}_k$ with either $\vect{R}_{\mathsf{V},k}\otimes\vect{R}_{\mathsf{H},k}$ (\ac{KBA} curves) or $\vect{R}^\star_{\mathsf{V},k} \otimes \vect{R}^\star_{\mathsf{H},k}$ (\ac{NKP} curves). 
We see that \ac{KBA} yields a lower $\mathsf{NSAE}_\mathbf{A}$, which results in a superior estimation accuracy when implementing \eqref{eq:A-generic}, as shown in the next section. 

\subsection{\acs{NMSE} evaluation}\label{perfect_upa:performance}

\begin{figure}[t]
  \begin{center}
    {\includegraphics[width=\columnwidth]{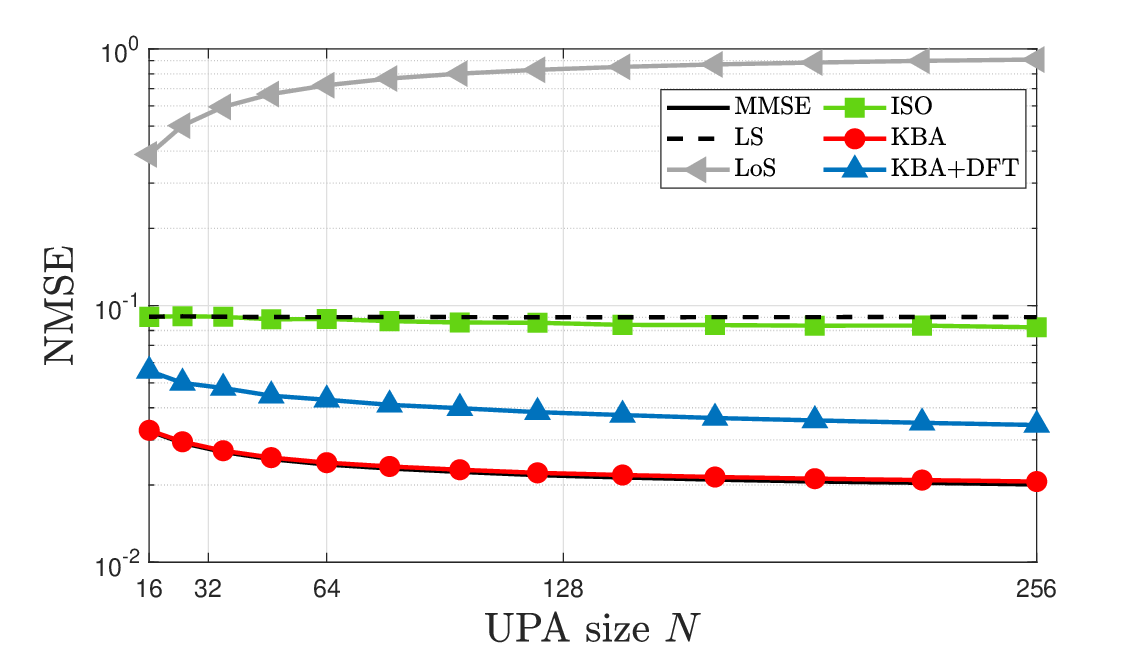}}
    \caption{\acs{NMSE} for a square \ac{UPA} with $N_\mathsf{H}=N_\mathsf{V}=\sqrt{N}$. The UPA parameters are reported in \tablename~\ref{tab:UPA}. The simulation parameters are those of \tablename~\ref{tab:system}.}
    \label{fig03}
  \end{center}
\end{figure}

Consider a square \ac{UPA} with $N_\mathsf{H}=N_\mathsf{V}=\sqrt{N}$ and all the other relevant parameters listed in \tablename~\ref{tab:UPA}. \figurename~\ref{fig03} shows the \ac{NMSE} in \eqref{eq:mse} as a function of $N=N_\mathsf{H}N_\mathsf{V}$. The matrix $\vect{A}_k$ in \eqref{eq:mse} depends on the specific channel estimation scheme. We consider \ac{MMSE}, \ac{LS}, \ac{LoS}, \ac{ISO}, and the approximate \ac{MMSE} based on \ac{KBA}.
We evaluate the average \ac{NMSE} for a \ac{UE} randomly placed in the simulation setup detailed in \tablename~\ref{tab:system}, in which the received \ac{SNR} is $\beta_k\tau_p\rho/\sigma^2$. The path loss $\beta_k$ is computed following \cite[\sectionname~2]{massivemimobook}, i.e., 
\begin{align}\beta_k=                        -148.1\dB - 37.6\log_{10}\left(\frac{d_k}{1 \km}\right)\qquad\textrm{[dB]}
\end{align}
where $d_k$ is the distance of \ac{UE} $k$ from the \ac{BS}.
It is worth noting that, in the considered network setup, in which the \acp{UE} are uniformly randomly placed with distances in $[\underline{d}, \overline{d}]$, the \ac{pdf} of the elevation angle $\overline{\theta}$ is 
\begin{align}\label{eq:pdf_theta}
  f_{\overline{\Theta}}\left(\theta\right)=\frac{2b^2}{(\overline{d}^2-\underline{d}^2) \sin^2\theta \left|\tan\theta\right|},
\end{align}
where $b$ is the \ac{BS} height on the azimuth plane. Accordingly, small values of $\left|\theta\right|$ are much more likely to occur than large values.

The results of \figurename~\ref{fig03} show that the accuracy of the \ac{KBA}-based estimator attains that of the optimal \ac{MMSE}, irrespective of the size $N$. This happens despite the error in the approximation of $\vect{R}_k$, because, as can be deduced from the numerical values in \figurename~\ref{fig02} and \eqref{eq:pdf_theta}, the \acp{NSAE} \eqref{eq:nsae} and \eqref{eq:nsae_a} are small enough to guarantee that $\vect{A}_k^\mathsf{Kron}\vect{y}_k$ yields approximately the same \ac{NMSE} \eqref{eq:mse} as $\vect{A}_k^\mathsf{MMSE}\vect{y}_k$. The accuracy of the \ac{NKP}-based estimator, not reported in \figurename~\ref{fig03}, overlaps both the \ac{MMSE} and the \ac{KBA} ones for the same reason. Similar results can be obtained with a \ac{UPA} with a rectangular shape. 

\begin{table}[t!]
\renewcommand{\arraystretch}{1.5}
\centering
\caption{\acs{UPA} parameters.}
\begin{tabular}{l|c}
\hline
\textbf{Parameter} & \textbf{Value} \\
\hline
Carrier frequency                    & $f_0 = 3\GHz$\\
Wavelength               & $\lambda=10\cm$\\
Vertical inter-element spacing & 
  $\Delta_\mathsf{V} = \lambda/2=5\cm$\\
Horizontal inter-element spacing & 
  $\Delta_\mathsf{H} = \lambda/2=5\cm$\\
Height            & $b = 10\meter$\\
\hline
\end{tabular}
\label{tab:UPA}
\end{table}

\begin{table}[t!]
\renewcommand{\arraystretch}{1.5}
\centering
\caption{Simulation parameters.}
\begin{tabular}{l|c}
\hline
\textbf{Parameter} & \textbf{Value} \\
\hline
Minimum distance from \acs{UPA} $\underline{d}$ & $5\meter$\\
Maximum distance from \acs{UPA} $\overline{d}$  & $100\meter$\\
Azimuth range $\varphi$                                   & 
  $\left[-60^\circ, +60^\circ\right]$\\
  Azimuth spreading $\sigma_\varphi$              & $10^\circ$\\
Elevation range $\theta$ 
& $\left[-63.4^\circ, -5.7^\circ\right]$\\
Elevation spreading $\sigma_\theta$             & $10^\circ$\\
Azimuth/elevation scattering distribution       & Gaussian\tablefootnote{The same simulation setup using a Laplacian distribution provide very similar results, not reported here for the
sake of brevity.}\\
Path loss reference distance                    & $1\km$\\
Channel gain at reference distance              & $-148.1\dB$\\
Path loss exponent & $3.76$\\
Communication bandwidth $B$                     & $100\MHz$\\
Transmit power $\rho$                           & $20\dBm$\\
Noise power                            & $-87\dBm$\\
Length of pilot sequence $\tau_p$               & $10$\\
Length of coherence block $\tau_c$              & $200$\\
\hline
\end{tabular}
\label{tab:system}
\end{table}

\begin{figure}[t]
  \begin{center}
 {\includegraphics[width=\columnwidth]{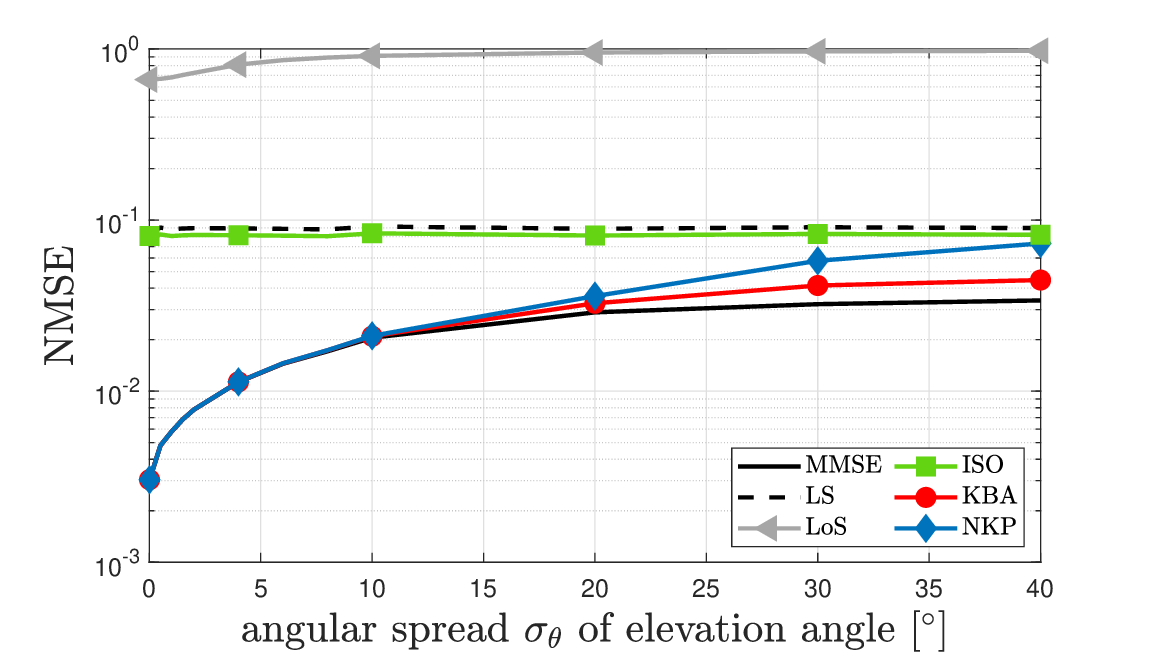}} 
    \caption{Average \acs{NMSE} as a function of the angular spread $\sigma_\theta$ of the elevation angle for the \acs{UPA} with parameters reported in \tablename~\ref{tab:UPA} and $N_\mathsf{H}=N_\mathsf{V}=16$. The simulation parameters are those of \tablename~\ref{tab:system}.}    \label{fig04}
  \end{center}
\end{figure}

The impact of the angular spread $\sigma_{\theta}$ is evaluated in \figurename~\ref{fig04}. The same simulation setup of \figurename~\ref{fig03} is considered. Also, we assume $N_\mathsf{H}=N_\mathsf{V}=16$ and  set $\sigma_\varphi=10^\circ$.
The results show that the \ac{KBA}-based estimator achieves the best accuracy, with a negligible gap with respect to the \ac{MMSE} estimator in the range of interest for typical cellular scenarios $\left[5^\circ, 15^\circ\right]$, as discussed at the end of \sectionname~\ref{sec:upa_system_model}.
For $10^\circ \le \sigma_\theta \le 40^\circ$, the difference between \ac{KBA} and \ac{NKP} increases, thus confirming the findings of \figurename~\ref{fig02}. A similar behavior can be observed using different values for the azimuth angular spread $\sigma_\varphi$.


\section{Reduced-complexity methods for \acs{ULA}}\label{sec:perfect_ula}

The \ac{UPA} model introduced in \sectionname~\ref{sec:upa_system_model} is general enough to encompass both vertical and horizontal configurations, obtained by either $N_\mathsf{H}=1$ or $N_\mathsf{V}=1$, respectively. In those cases, the \ac{KBA}-based approach has exactly the same complexity of the optimal \ac{MMSE} one. This is due to the fact that, when $N_\mathsf{H}$ (respectively, $N_\mathsf{V}$) is equal to $1$, based on \eqref{eq:R_H_Matlab} [respectively, \eqref{eq:R_V_Matlab}], the matrix $\vect{R}_{\mathsf{H},k}(\overline{\theta})$
[respectively, $\vect{R}_{\mathsf{V},k}(\overline{\theta})$] is \emph{exactly} the scalar $1$, and hence 
$\vect{R}_k=\vect{R}_{\mathsf{V},k}(\overline{\theta})$ [respectively, $\vect{R}_k=\vect{R}_{\mathsf{H},k}(\overline{\theta})$]. For this reason,
not only the complexity, but also the channel estimation performance of the \ac{KBA}-based method coincides with the \ac{MMSE} one.


In this section, we develop a channel estimation scheme for a horizontal \ac{ULA} (i.e., $N=N_\mathsf{H}$) that exploits the correlation 
induced by the array geometry and propagation conditions to approach \ac{MMSE} performance, while having a computational complexity that 
scales log-linearly with $N$. The adaptation to the vertical \ac{ULA} configuration (with $N=N_\mathsf{V}$) is straightforward and
not reported here for the sake of brevity.

\subsection{\acs{DFT}-based approximation}\label{perfect_ula:dft}

If a \ac{ULA} is used, the covariance
matrix  $\vect{R}_k$ is Hermitian Toeplitz, and it can be approximated with a suitable circulant matrix
$\vect{C}_k$ \cite{Pearl1971,Pearl1973,Wakin2017}, whose first row $\vect{c}_{k}=[c_k(0),c_k(1),\cdots,c_k(N-1)]$ is related to the first
row $\vect{r}_{k}=[r_k(0),r_k(1),\cdots,r_k(N-1)]$ of $\vect{R}_k$ by \cite{Wakin2017}
\begin{align}\label{eq:Ck}
  c_k(n)=\begin{cases}
  r_k(0)& \text{$n=0$}, \\
  \dfrac{(N-n) r_k(n)+n r^{*}_k (N-n)}{N} & \text{$n = 1,\ldots,N-1$}.
  \end{cases}
\end{align}
Any circulant matrix can be unitarily diagonalized using the \ac{DFT} matrix, i.e., 
$ {\vect{C}}_k={\vect{F}} {\vect{\Lambda}}_{k}  {\vect{F}}^{\Htran}$
where ${\vect{F}}=[{\vect{f}}_{0} \, {\vect{f}}_{1} \, \cdots \, {\vect{f}}_{N-1}]$ is the inverse \ac{DFT} matrix, with
$[{\vect{f}}_n]_m=\frac{1}{\sqrt{N}}e^{\imagunit 2 \pi mn/N}$ for $0 \le m,n \le N-1$, and ${\vect{\Lambda}}_{k}$ is 
the diagonal matrix containing the eigenvalues of ${\vect{C}}_k$, i.e.,
\begin{align}\label{eq:DFT_firstrow}
  [{\vect{\Lambda}}_k]_{n,n}=\sum\limits_{m=0}^{N-1}c_{k}(m)e^{-\imagunit 2 \pi mn/N}  
\end{align}
which are obtained by taking the \ac{DFT} of the first row of $\vect{C}_k$. Replacing $\vect{R}_k$ with $\vect{C}_k$ into \eqref{eq:A-MMSE} yields $\widehat{\vect{h}}_k^{\mathsf{DFT}} = {\bf A}_k^{\mathsf{DFT}}\vect{y}_k$,
with
\begin{align}\label{eq:A-DFT}
  \vect{A}_k^{\mathsf{DFT}} = \frac{1}{\tau_p \sqrt{\rho}} {\vect{F}} \, {\vect{\Lambda}_k} 
  \left({\vect{\Lambda}_k} +\frac{1}{\gamma}{\vect{I}}_N \right)^{-1}{\vect{F}}^{\Htran}.
\end{align}
We call it the \emph{\ac{DFT}-based channel estimator}. Its complexity derives from the pre-computation phase, which 
is $\mathcal{O}(N \log N)$ due to the computation of ${\vect{\Lambda}}_k$ through \eqref{eq:DFT_firstrow}, and from the 
computation of the matrix-vector product, which is again $\mathcal{O}(N \log N)$, since the \ac{DFT} matrix $\vect{F}$ 
and its inverse are involved. Hence, the  complexity of the \ac{DFT}-based estimator is $\mathcal{C}_{\mathsf{DFT}}=\mathcal{O}(N \log N)$. 
Note that, similarly to the \ac{KBA} case, the \ac{DFT}-based estimator depends on the true covariance matrix ${\vect R}_k$, which 
must be estimated as with the \ac{MMSE} estimator.

\begin{figure}[t]
  \begin{center}
    {\includegraphics[width=\columnwidth]{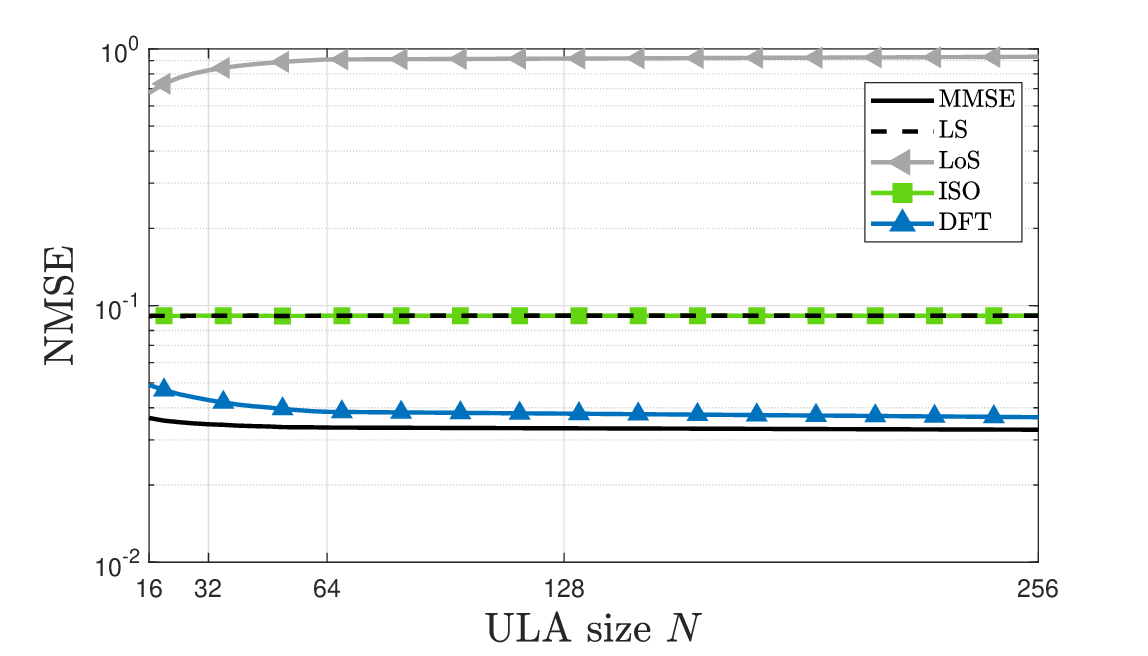}}
    \caption{\acs{NMSE} as a function of the \acs{ULA} size $N$.}
    \label{fig05}
  \end{center}
\end{figure}

\subsection{NMSE analysis}\label{perfect_ula:performance}

Consider a horizontal \ac{ULA}, using the relevant parameters listed in \tablesname~\ref{tab:UPA} and \ref{tab:system}. \figurename~\ref{fig05} shows the \ac{NMSE} in \eqref{eq:mse} as a function of $N$. We see that the accuracy of the \ac{DFT}-based estimator is comparable with the \ac{MMSE}, and the difference
decreases as $N$ increases. This is due to the fact that the circulant approximation $\vect{C}_{k}$ 
of the covariance matrix $\vect{R}_{k}$ improves as $N$ grows.
Interestingly, the circulant approximation is already quite tight for $N=64$ (which means $L=N \Delta_\mathsf{H}=3.2\meter$). 
More importantly, this is obtained with a complexity of $\mathcal{O}(N \log N)$, instead of $\mathcal{O}(N^3)$. 
If $N=64$, this corresponds to two orders of magnitude of computational saving compared to \ac{MMSE}. Please refer to \cite{Damico_2023} for numerical results of vertical \acp{ULA} in terms of performance as a function of the \ac{ULA} size and of the angular spreads.

\subsection{The failure of the combination with \ac{KBA}}\label{perfect_ula:kbadft}

Inspired by the circulant approximation that exploits
the Toeplitz structure of $\vect{R}_k$ in a \ac{ULA}, one might be tempted to adopt the same approach for the matrices $\vect{R}_{\mathsf{H},k}$ and $\vect{R}_{\mathsf{V},k}$ in \eqref{eq:KBA} since they are both Hermitian Toeplitz. This means applying \eqref{eq:Ck} to the first row of either matrix $\vect{R}_{\mathsf{H},k}$ or $\vect{R}_{\mathsf{V},k}$.

Unfortunately, while the approximation errors for each Hermitian Toeplitz matrix, which can be relevant especially for the off-diagonal elements, appear to be tolerable when implementing the \ac{DFT}-based scheme \eqref{eq:A-DFT}, this is not true when the approximated versions of the matrices $\vect{R}_{\mathsf{H},k}$ or $\vect{R}_{\mathsf{V},k}$ are re-combined. In fact, when  we apply the Kronecker product to implement the \ac{KBA}-based approach, the approximation errors get amplified.

\begin{figure}[t]
\begin{center}
\includegraphics[width=\columnwidth]{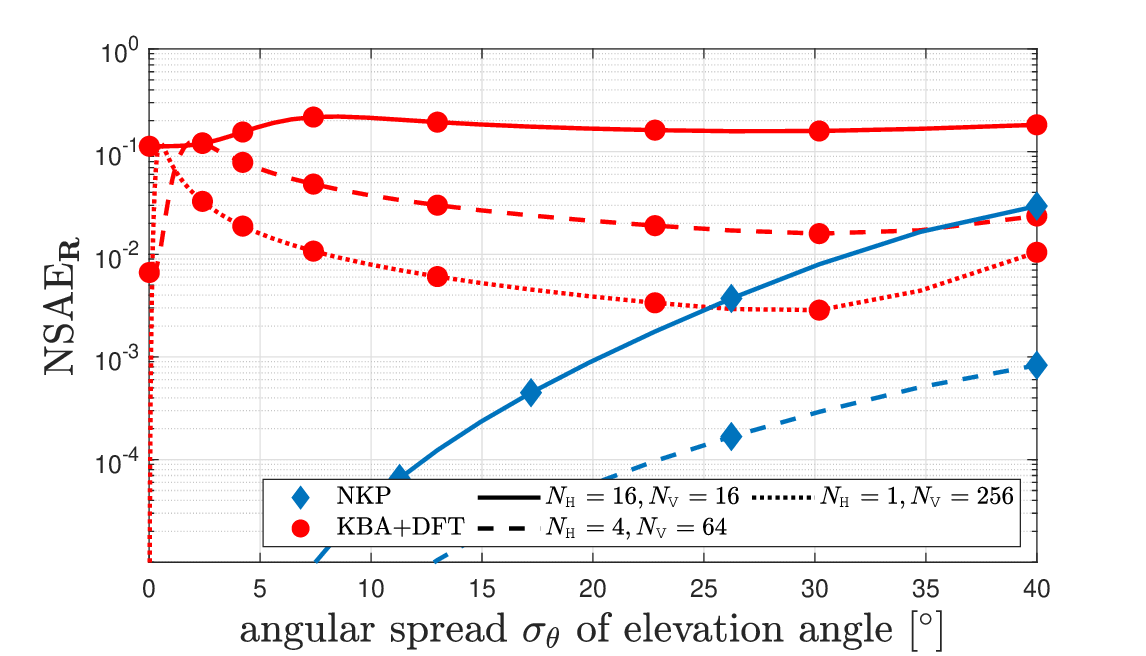}
\caption{\acs{NSAE} of the correlation matrix as a function of the elevation spreading for the combined \acs{KBA}/\acs{DFT} approach and the \ac{NKP} solution.}
\label{fig06}
\end{center}
\end{figure}

To showcase this, \figurename~\ref{fig06} illustrates $\mathsf{NSAE}_\mathbf{R}$ as a function of $\sigma_\theta$ for the following configurations: $(N_\mathsf{H}=N_\mathsf{V}=16)$ (solid lines), $(N_\mathsf{H}=1, N_\mathsf{V}=256)$ (dashed lines), and $(N_\mathsf{H}=4, N_\mathsf{V}=64)$ (dotted lines). We consider $\overline{\varphi}=\overline{\theta}=0^\circ$, and the scattering of both angles is modeled as Gaussian, with $\sigma_\theta=\sigma_\varphi=10^\circ$. Red lines with circular markers represent the results by combining the \ac{KBA} and \ac{DFT} approaches described above, whereas blue lines with diamond markers report the \ac{NKP} solution introduced in \sectionname~\ref{perfect_upa:kba}.\footnote{The curve for the \ac{ULA} case $(N_\mathsf{H}=1, N_\mathsf{V}=256)$ is not reported, as the \ac{NKP} solution coincides with $\vect{R}_k$, and hence $\mathsf{NSAE}=0$.} As can be seen, keeping the \ac{UPA} size $N$ constant, the performance gap for all configurations is significant, especially for moderate values of $\sigma_\theta$, which are of particular interest in practical working conditions. The behavior illustrated in \figurename~\ref{fig06} can be replicated with any set of parameters $(N_\mathsf{H}>1,N_\mathsf{V}>1)$.

The impact in terms of channel estimation accuracy can be seen in \figurename~\ref{fig03} (blue curve), which confirms the poor performance achieved by the \ac{KBA}/\ac{DFT} combined 
method (the accuracy is significantly worse than that achieved by the \ac{KBA} estimator).

\section{Channel estimation with imperfect knowledge of covariance matrices}\label{sec:imperfect_knowledge}

So far, we have assumed perfect knowledge of $\vect{R}_{k}$. This may not be the case in practical scenarios since $\vect{R}_{k}$ changes for various reasons \cite{sanguinettiTCOM2020}. Measurements suggest that these changes are orders of magnitude slower than the fast variations of the channel \cite{sanguinettiTCOM2020}. Therefore, it is reasonable to assume that they do not change over $\tau_s=B_sT_s/\tau_c$ blocks \cite{sanguinettiTCOM2020}, where $B_s$ is the system bandwidth, $T_s$ is the timing interval over which the second-order statistics remain unaltered, and $\tau_c$ is the length (in samples) of the channel coherence block, that equals the product between the coherence bandwidth and the coherence time of the channel. To quantify $\tau_s$, let us consider a mobile scenario using $B_s=10\MHz$ and $T_s=0.5\seconds$  \cite{Bjornson2016}. Using $\tau_c=200$, as considered throughout the paper, which is compatible with a coherence bandwidth of $200\kHz$ and a coherence time of $1\ms$, we get $\tau_s=25000$.

Suppose the \ac{BS} has the pilot symbols $\vect{\phi}_k$ in $M \le \tau_s$ coherence
blocks. We denote the corresponding observations by $\vect{y}_k[1],\ldots,\vect{y}_k[M]$. An estimate of ${\vect{Q}}_{k}$ 
can be obtained by computing the sample correlation matrix given by
\begin{align}\label{eq:Qsample}
  \widehat{\vect{Q}}_{k}^{\mathsf{sample}} = \frac{1}{M} \sum_{m=1}^{M} \frac{\vect{y}_k[m] \vect{y}_k^{\Htran}[m]}{\rho \tau_p^2}.
\end{align}
The computation of $\widehat{\vect{Q}}_{k}^{\mathsf{sample}}$ requires $\mathcal{O}(MN^2)$ operations, as it involves $M N^2$ complex multiplications.

A better estimation is typically obtained through matrix regularization by computing the convex combination \cite{sanguinettiTCOM2020}:
\begin{align}\label{eq:regularization}
  \widehat{\vect{Q}}_{k}(\eta) = \eta\widehat{\vect{Q}}_{k}^{\mathsf{sample}}+ (1-\eta) \widehat{\vect{Q}}_{k}^{\mathsf{diag}}, \quad  \eta \in [0,1]
\end{align}
where $\widehat{\vect{Q}}_{k}^{\mathsf{diag}}$ contains the main diagonal of $\widehat{\vect{Q}}_{k}^{\mathsf{sample}}$.
The regularization makes $\widehat{\vect{Q}}_{k}(\eta)$ a full-rank matrix for any $\eta \in [0,1)$, and $\eta$ can be tuned
(for example by using numerical methods) to purposely underestimate the off-diagonal elements when these are considered unreliable. 
Once $\widehat{\vect{Q}}_{k}(\eta)$ is computed, an estimate of ${\vect{R}}_k$ follows:
\begin{align}\label{eq:Rsample}
  \widehat{\vect{R}}_k(\eta) = \widehat{\vect{Q}}_{k}(\eta) - \frac{1}{\gamma}{\vect{I}}_N
\end{align}
which requires only knowledge of $\gamma$, i.e., the \ac{SNR} during the pilot transmission phase.

\subsection{Improved estimation of ${\vect{Q}}_k$ and  ${\vect{R}}_k$}\label{imperfect_knowledge:improved}

We now develop an improved estimation scheme of the correlation matrix $\vect{Q}_k$ that can be used with \acp{UPA}. 
For the ease of notation, we will drop the subscript $k$ from now on. 

To accomplish this task, we notice that $\vect{Q}$, analogously to $\vect{R}$, is Hermitian \emph{block}-Toeplitz (and not simply Toeplitz), with $N_\mathsf{V} \times N_\mathsf{V}$ blocks with size $N_\mathsf{H}\times N_\mathsf{H}$, i.e.,
\begin{align}\label{eq:Q_blockToeplitz}
  \vect{Q}=\begin{bmatrix}
    \vect{Q}_{1,1} & \vect{Q}_{1,2} & \vect{Q}_{1,3} & \ldots & \vect{Q}_{1,N_\mathsf{V}}\\
    \vect{Q}_{1,2}^{\Htran} & \vect{Q}_{1,1} & \vect{Q}_{1,2} & \ldots & \vect{Q}_{1,N_\mathsf{V}-1}\\
    \vect{Q}_{1,3}^{\Htran} & \vect{Q}_{1,2}^{\Htran} & \vect{Q}_{1,1} & \ldots & \vect{Q}_{1,N_\mathsf{V}-2}\\
    \vdots & \vdots & \vdots & \ddots & \vdots\\
    \vect{Q}_{1,N_\mathsf{V}}^{\Htran} & 
    \vect{Q}_{1,N_\mathsf{V}-1}^{\Htran} & 
    \vect{Q}_{1,N_\mathsf{V}-2}^{\Htran} & 
    \ldots & \vect{Q}_{1,1}
  \end{bmatrix},
\end{align}
where $\vect{Q}_{1,j}$ denotes the $N_\mathsf{H}\times N_\mathsf{H}$ block obtained by $\vect{Q}$ as 
(using \textsc{MatLab} colon notation)
\begin{align}\label{eq:Q_singleBlock}
  \vect{Q}_{1,j}=\left[\vect{Q}\right]_{1:1:N_\mathsf{H}, 
  \left(j-1\right)N_\mathsf{H}+1:1:j N_\mathsf{H}}.
\end{align}
This means that, due to the block-Toeplitz structure, 
\begin{align}\label{eq:Q_singleBlock_bt}
  \vect{Q}_{1+m,j+m}=\vect{Q}_{1,j}
\end{align}
for $j=1,\dots,N_\mathsf{V}-1$ and $m=1,\dots,N_\mathsf{V}-j$, and that 
\begin{align}\label{eq:Q_singleBlock_h}
  \vect{Q}_{j,1}=\vect{Q}_{1,j}^{\Htran},
\end{align}
because of the Hermitian symmetry of the covariance matrix.

To proceed with the estimation, we denote by $\widehat{\vect{Q}}_{1,j}$ the estimate of $\vect{Q}_{1,j}$ 
that takes the block-Toeplitz structure \eqref{eq:Q_blockToeplitz} into account, by averaging the blocks of $\widehat{\vect{Q}}^{\mathsf{sample}}$ over the same (block) diagonal, i.e.:
\begin{align}\label{eq:Q_singleBlock_estimate}
  \widehat{\vect{Q}}_{1,j}=\frac{1}{N_\mathsf{V}-j+1}
  \sum_{m=1}^{N_\mathsf{V}-j+1}{\widehat{\vect{Q}}^{\mathsf{sample}}_{m,j+m-1}},
\end{align}
for $j=1,\dots,N_\mathsf{V}$, where ${\widehat{\vect{Q}}^{\mathsf{sample}}_{m,j+m-1}}$ is the $N_\mathsf{H}\times N_\mathsf{H}$ block corresponding to the rows $(m-1)N_\mathsf{H}+1,\ldots,mN_\mathsf{H}-1$ and to the columns $(j+m-2)N_\mathsf{H}+1,\ldots,(j+m-1)N_\mathsf{H}-1$ of $\widehat{\vect{Q}}^{\mathsf{sample}}$ in \eqref{eq:Qsample}.


It is worth noting that $\vect{Q}_{1,j}$ is an $N_\mathsf{H}\times N_\mathsf{H}$ Toeplitz (but \emph{not} Hermitian) matrix, and we can thus improve the estimation further, by adapting the matrix-based criterion \eqref{eq:Q_singleBlock_estimate} to 
the elements of $\vect{Q}_{1,j}$. In particular, the estimate
${\widehat{\vect{Q}}^{\mathsf{toe}}_{1,j}}$ can be obtained as follows: the first row is computed as
\begin{align}\label{eq:Qtoe}
  \left[\widehat{\vect{Q}}^{\mathsf{toe}}_{1,j}\right]_{1,\ell}=
  \frac{1}{N_\mathsf{H}-\ell+1}\sum_{m=1}^{N_\mathsf{H}-\ell+1}
  {\left[\widehat{\vect{Q}}_{1,j}\right]_{m,\ell+m-1}}
\end{align}
for $\ell=1,\dots,N_\mathsf{H}$, whereas the first column can be computed as
\begin{align}\label{eq:Qtoe2}
  \left[\widehat{\vect{Q}}^{\mathsf{toe}}_{1,j}\right]_{\ell,1}=
  \frac{1}{N_\mathsf{H}-\ell+1}\sum_{m=1}^{N_\mathsf{H}-\ell+1}
  {\left[\widehat{\vect{Q}}_{1,j}\right]_{\ell+m-1,m}}
\end{align}
for $\ell=1,\dots,N_\mathsf{H}$. Exploiting the Toeplitz structure, the remaining elements of $\widehat{\vect{Q}}^{\mathsf{toe}}_{i,j}$ can be found as
\begin{align}\label{eq:Qtoe3}
  \left[\widehat{\vect{Q}}^{\mathsf{toe}}_{1,j}\right]_{1+m,\ell+m}&=
  \left[\widehat{\vect{Q}}^{\mathsf{toe}}_{1,j}\right]_{1,\ell}\\
  \left[\widehat{\vect{Q}}^{\mathsf{toe}}_{1,j}\right]_{\ell+m,1+m}&=
  \left[\widehat{\vect{Q}}^{\mathsf{toe}}_{1,j}\right]_{\ell,1}
\end{align}
for $\ell=1,\dots,N_\mathsf{H}-1$ and $m=1,\dots,N_\mathsf{H}-\ell$.

\begin{figure}[t]
  \begin{center}
    {\includegraphics[width=\columnwidth]{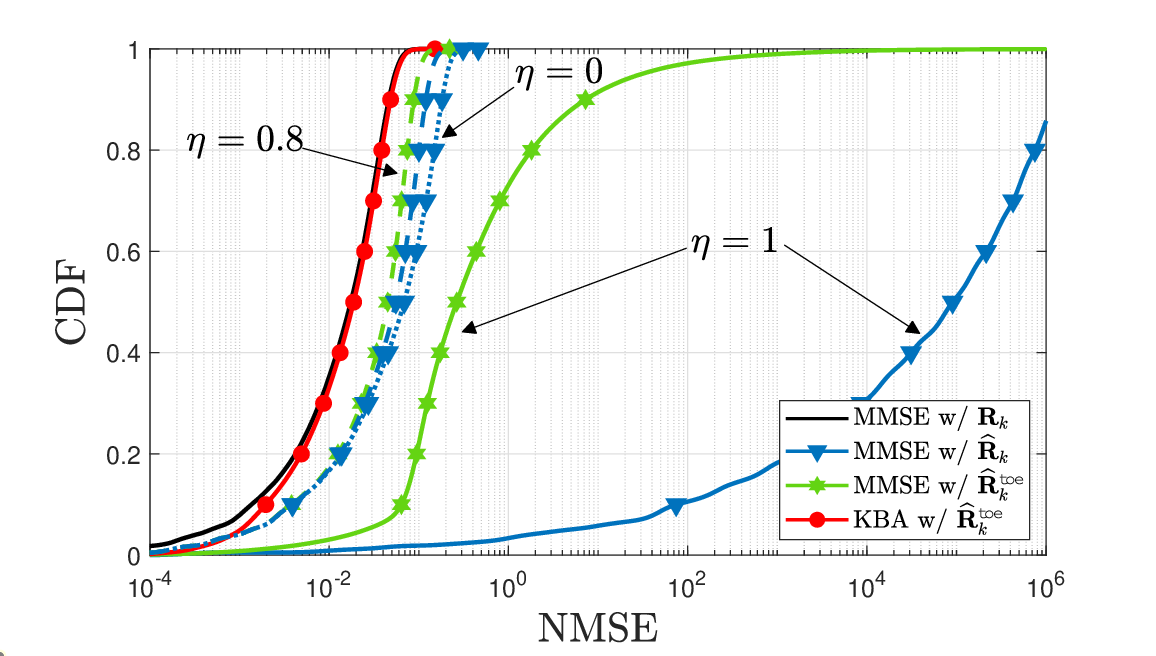}}
    \caption{\acs{CDF} of the \acs{NMSE} as a function of the estimation technique using $M=50$ (black: \acs{MMSE} with perfect knowledge of $\vect{R}_k$; red: \acs{KBA} using $\widehat{\vect{R}}^{\mathsf{toe}}_k$ \eqref{eq:Rtoe}; blue: \acs{MMSE} using $\widehat{\vect{R}}_k(\eta)$ \eqref{eq:Rsample} in \eqref{eq:A-MMSE} as a function of $\eta$; green: \acs{MMSE} using $\widehat{\vect{R}}^{\mathsf{toe}}_k(\eta)$ \eqref{eq:Rtoe} in \eqref{eq:A-MMSE} as a function of $\eta$).}
    \label{fig07}
  \end{center}
\end{figure}

Once $\vect{Q}_{1,j}$ is available for $j=1,...,N_\mathsf{V}$, the estimate $\widehat{\vect{Q}}^{\mathsf{toe}}$ of the covariance matrix $\vect{Q}$ can be found by the same structure as in \eqref{eq:Q_blockToeplitz}, using 
$\widehat{\vect{Q}}^{\mathsf{toe}}_{1,j}$ instead of ${\vect{Q}}_{1,j}$. An estimate
of $\vect{R}$ is finally obtained as
\begin{align}\label{eq:Rtoe}
  \widehat{\vect{R}}^{\mathsf{toe}}=\widehat{\vect{Q}}^{\mathsf{toe}}-\frac{\vect{I}_N}{\gamma}.
\end{align}
The complexity of the estimator above is mainly due to the computation of $\widehat{\vect{Q}}^{\mathsf{sample}}$,
and thus is comparable to the one not exploiting the Toeplitz structure, as shown in \sectionname~\ref{sec:complexity}.

To implement the \ac{MMSE} estimation scheme, we can replace $\vect{R}$ and $\vect{Q}$ in \eqref{eq:A-MMSE} with
$\widehat{\vect{R}}^{\mathsf{toe}}$ and $\widehat{\vect{Q}}^{\mathsf{toe}}$, respectively.
We can also adopt the \ac{KBA}- and the \ac{DFT}-based methods,
for the \ac{UPA} and \ac{ULA} cases, respectively, by considering $\widehat{\vect{R}}^{\mathsf{toe}}$ in \eqref{eq:KBA} and \eqref{eq:Ck}, instead of $\vect{R}$.

\begin{figure}[t]
  \begin{center}
    {\includegraphics[width=\columnwidth]{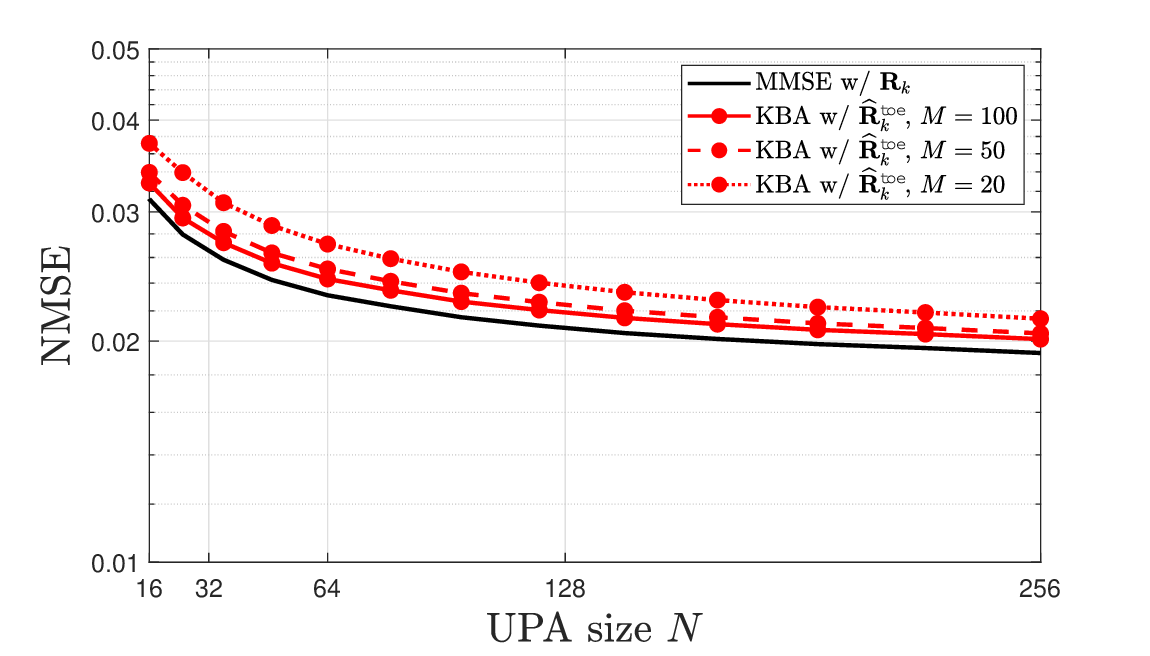}}
    \caption{\acs{NMSE} as a function of the size $N$ (square \acs{UPA}, with $N_\mathsf{H}=N_\mathsf{V}$) with imperfect statistical knowledge (\acs{KBA}-based estimator).}
    \label{fig08}
  \end{center}
\end{figure}

\subsection{Performance analysis}\label{imperfect_upa:performance}

We now evaluate the accuracy of the different estimators, using the system setup detailed in \tablename~\ref{tab:system}, assuming a square \ac{UPA} with parameters listed in \tablename~\ref{tab:UPA}. \figurename~\ref{fig07} reports the \ac{CDF} of \ac{NMSE} in \eqref{eq:mse} asssuming that $M=50$ observations are used for the estimation of the channel covariance matrix. In particular, the red line with circular markers corresponds to the \ac{KBA} estimator that uses $\widehat{\vect{R}}^\mathsf{toe}_k$ in \eqref{eq:Rtoe}, whereas blue and green lines correspond to the \ac{MMSE} estimator when using \eqref{eq:Rsample} and \eqref{eq:Rtoe}, respectively. Solid, dashed, and dotted lines report the results obtained with the regularization parameter $\eta$ introduced in \eqref{eq:regularization} with values $\{1, 0.8, 0\}$, respectively.\footnote{In the case of the \ac{MMSE} estimator using \eqref{eq:Rtoe}, the same regularization method \eqref{eq:regularization} is applied to the matrix $\widehat{\vect{Q}}^{\mathsf{toe}}$. Note that, for $\eta=0$, the two methods coincide, and only the lower triangles is reported.} For comparison, we also report the \ac{NMSE} achieved by the \ac{MMSE} estimator with perfect knowledge of $\vect{R}_k$ (black line). 

The \ac{KBA} estimator performs close to the \ac{MMSE} estimator with perfect knowledge of channel statistics, whereas the \ac{MMSE} estimator with imperfect knowledge of channel statistics shows a gap that depends on $\eta$, and thus confirming the benefits yielded by the regularization technique \eqref{eq:regularization}. Note also that the \ac{MMSE} estimator using \eqref{eq:Rtoe} outperforms the one using \eqref{eq:Rsample}. For this reason, in the remainder of the paper, we will adopt \eqref{eq:Rtoe} with a regularization factor $\eta=0.8$.\footnote{The optimal $\eta$ does depend on the network setup and the \ac{UPA} parameters, including size and spacing.} Additional simulations, not reported for the sake of clarity, show that: \emph{i})~regularizing the \ac{KBA} estimator does \emph{not} provide additional benefits, as the performance is already tight to the optimal one; and \emph{ii})~using \eqref{eq:Rsample} instead of \eqref{eq:Rtoe} produces poorer performance of the \ac{KBA} estimator. Hence, in the remainder of the paper we will use \eqref{eq:Rtoe} when considering \ac{KBA}.

\figurename~\ref{fig08} reports the average as a function of the size $N$ for a square \ac{UPA}, obtained by averaging over all possible \ac{UE} positions in the system setup detailed in \tablename~\ref{tab:system}.
As can be seen, an estimation accuracy comparable to that obtained with perfect knowledge of $\vect{R}_{k}$ is already achieved with $M=20$. Similar trends are observed for different array sizes and/or scattering scenarios. The results concerning \ac{ULA} configurations are reported in \cite{Damico_2023}.

\section{Complexity analysis}\label{sec:complexity}
Next, we assess the computational burden in terms of complex operations (additions and/or multiplications), required by the \emph{entire} channel estimation procedure of the investigated estimators. 
Without loss of generality, next we consider $N_\mathsf{H}\ge N_\mathsf{V}$. The analysis can be simply extended to the case $N_\mathsf{H}<N_\mathsf{V}$.

\subsection{Computation of ${\widehat{\vect{Q}}_k^{\mathsf{toe}}}$}\label{complexity:Q}



From \eqref{eq:Q_singleBlock_estimate}, the calculation of $\widehat{\vect{Q}}_{1,j}$, for $j=1,\dots,N_\mathsf{V}$, requires a total of approximately $N^2/2$ additions and $N_\mathsf{H}N$ multiplications. Analogously, the evaluation of ${\widehat{\vect{Q}}^{\mathsf{toe}}_{1,j}}$ in \eqref{eq:Qtoe}, with $j=1,\dots,N_\mathsf{V}$, requires a total of $NN_\mathsf{H}/2$ additions and $N$ multiplications. Thanks to Toeplitz-block Toeplitz structure, the computation of ${\widehat{\vect{Q}}_k^{\mathsf{toe}}}$ has a complexity order of $\mathcal{O}(MN^2 + N^2/2 + N_\mathsf{H}N) \approx\mathcal{O}(MN^2)$. No additional complexity is incurred compared to the unstructured $\widehat{\vect{Q}}_{k}^{\mathsf{sample}}$.

\subsection{Computation of matrix $\vect{A}_{k}$ via ${\widehat{\vect{R}}_k^{\mathsf{toe}}}$ and ${\widehat{\vect{Q}}_k^{\mathsf{toe}}}$}\label{complexity:A}

The \ac{MMSE} estimator is characterized by the matrix 
\begin{align}\label{eq:A-MMSE_Est}
\widehat{\vect{A}}_k^{\mathsf{MMSE}}= \frac{1}{\tau_p \sqrt{\rho}}{\widehat{\vect{R}}}^{\mathsf{toe}}_k \left(\widehat{\vect{Q}}_{k}^{\mathsf{toe}}\right)^{-1}
\end{align}
obtained from \eqref{eq:A-MMSE} by replacing the true covariance matrix $\vect{R}_k$ with its estimate $\widehat{\vect{R}}^{\mathsf{toe}}_k$. Once ${\widehat{\vect{Q}}_{k}^{\mathsf{toe}}}$ is available, $\widehat{\vect{R}}^{\mathsf{toe}}_k$ can easily be obtained through \eqref{eq:Rtoe} with $N$ additions. As for the computation of $(\widehat{\vect{Q}}^{\mathsf{toe}}_k)^{-1}$, an efficient algorithm for the inversion of a Toeplitz-block Toeplitz matrix can be found in \cite{WaxKailath1983}. It has a complexity $\mathcal{O}(N^2N_\mathsf{V})$. The maximum complexity occurs for square \acp{UPA}, where $N_\mathsf{H}=N_\mathsf{V}=\sqrt{N}$, resulting in $\mathcal{O}(N^{2.5})$.

\begin{table}[t]
\renewcommand{\arraystretch}{1.5}
\centering
\caption{Complexity of channel estimation schemes.}
\begin{tabular}{c|c|c|c}

\textbf{Scheme} & ${\widehat{\vect{Q}}^{\mathsf{toe}}}$ & $\vect{A}_k$ & $\vect{A}_k\vect{y}_k$ \\
\hline
\acs{LS}   & --                 & --                               & $\mathcal{O}(N)$ \\
\acs{LoS}  & --                 & --                               & $\mathcal{O}(N)$ \\
\acs{ISO}  & --                 & --                               & $\mathcal{O}(N^2)$ \\
\acs{MMSE} & $\mathcal{O}(MN^2)$ & $\mathcal{O}(N^2 N_{\mathsf V})$ & $\mathcal{O}(N^{2})$ \\
\acs{KBA}  & $\mathcal{O}(MN^2)$ & $\mathcal{O}(N_{\mathsf H}^3)$          & $\mathcal{O}((N_\mathsf{H}+N_\mathsf{V})N)$ \\
\acs{DFT}  & $\mathcal{O}(N^2)$ & $\mathcal{O}(N \log N)$          & $\mathcal{O}(N \log N)$ \\
\hline
\end{tabular}
\label{tab:complexity}
\end{table}

The \ac{KBA} estimator is obtained by approximating $\widehat{\vect{R}}_k^{\mathsf{toe}}$ with the Kronecker product $\widehat{\vect{R}}^{\mathsf{toe}}_{\mathsf{V},k} \otimes \widehat{\vect{R}}^{\mathsf{toe}}_{\mathsf{H},k}$. The matrix $\widehat{\vect{A}}_k^\mathsf{KBA}$ is derived from \eqref{eq:A-Kron} by replacing
\begin{align}
{\vect{K}}_{1}&=\vect{U}_{1} \vect{\Lambda}_{1}\vect{U}_{1}^{\Htran} \\{\vect{K}}_{2}&=\vect{U}_{2} \vect{\Lambda}_{2}\vect{U}_{2}^{\Htran} 
\end{align}
with 
\begin{align}\label{R_Vtoe}
\widehat{\vect{R}}^{\mathsf{toe}}_{\mathsf{V},k}&=\widehat{\vect{U}}_{\mathsf{V},k}^{\mathsf{toe}} \widehat{\vect{\Lambda}}_{\mathsf{V},k}^{\mathsf{toe}}(\widehat{\vect{U}}_{\mathsf{V},k}^{\mathsf{toe}})^{\Htran} \\ \label{R_Htoe}\widehat{\vect{R}}^{\mathsf{toe}}_{\mathsf{H},k}&= \widehat{\vect{U}}_{\mathsf{H},k}^{\mathsf{toe}} \widehat{\vect{\Lambda}}_{\mathsf{H},k}^{\mathsf{toe}}(\widehat{\vect{U}}_{\mathsf{H},k}^{\mathsf{toe}})^{\Htran}
\end{align}
respectively. By utilizing the algorithm proposed in \cite{Trench1989}, the computation of \eqref{R_Vtoe} and \eqref{R_Htoe} has a complexity of $\mathcal{O}(N_\mathsf{H}^3)$ and $\mathcal{O}(N_\mathsf{V}^3)$, respectively. Consequently, the factorization of $\widehat{\vect{A}}_k^\mathsf{KBA}$ can be achieved with a complexity of $\mathcal{O}(N_{\mathsf H}^3)$, which amounts to $\mathcal{O}(N^{1.5})$ for square \acp{UPA}.



\subsection{Computation of channel estimate $\widehat{\vect{h}}_k$}\label{complexity:h} 
All estimation schemes are \emph{linear} of the form given in \eqref{eq:A-generic}, based on $\vect{y}_k$. The complexity associated with the computation of $\vect{y}_k$ is $\mathcal{O}(N)$, since it requires $(\tau_p -1)N$ complex sums and $\tau_p N$ complex products.

The \ac{LS} estimator simply multiplies $\vect{y}_k$ and the real diagonal matrix in \eqref{eq:A-LS}. Accordingly, it has an overall complexity $\mathcal{O}(N)$. The \ac{MMSE} estimator multiplies $\vect{y}_k$ by the matrix \eqref{eq:A-MMSE_Est}. 
From the Appendix, it follows that it exhibits a computational complexity of $\mathcal{O}(N^2)$. This is because computing $(\widehat{\vect{Q}}_{k}^{\mathsf{toe}})^{-1} \vect{y}_k$ requires $N^2$ multiplications and (approximately) $N^2$ additions, and the same number of operations is needed to compute $\widehat{\vect{R}}_k^{\mathsf{toe}} (\widehat{\vect{Q}}_{k}^{\mathsf{toe}})^{-1} \vect{y}_k$.

The \ac{KBA} scheme makes use of the matrix $\widehat{\vect{A}}_k^\mathsf{KBA}$. 
From the Appendix, based on \eqref{eq:A-Kron} and exploiting the structure of \ac{KBA} due to the Kronecker products, the complexity associated with the computation of $\widehat{\vect{A}}_k^\mathsf{KBA} \vect{y}_k$ is $\mathcal{O}((N_\mathsf{H}+N_\mathsf{V})N)$. If $N_\mathsf{H}=N_\mathsf{V}=\sqrt{N}$, the complexity is $\mathcal{O}(N\sqrt{N})$.

\begin{figure}[t]
\begin{center}\vspace{-0.5cm}
\includegraphics[width=\columnwidth]{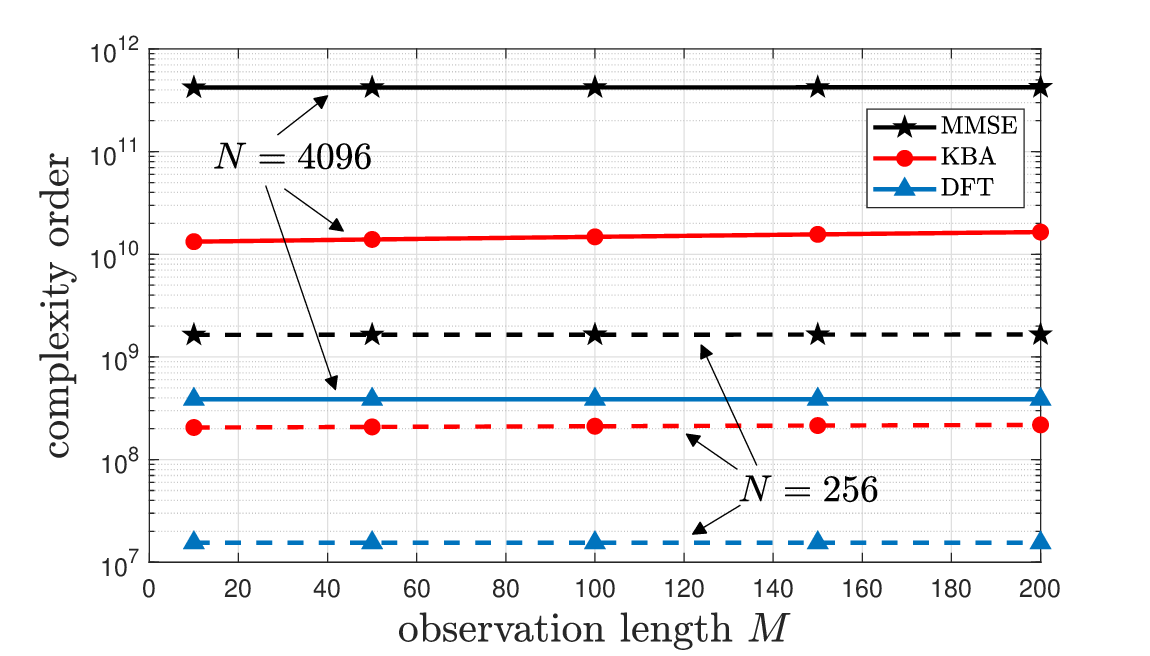}
\caption{Complexity order as a function of the observation length for different estimation schemes and array sizes ($\tau_s=25,000$).}
\label{fig:complexity}

\end{center}
\end{figure}

\subsection{Summary}

The above results are summarized in \tablename~\ref{tab:complexity}, which also contains the complexity for the \ac{DFT}-based method derived in \sectionname~\ref{sec:perfect_ula} (please refer to \cite{Damico_2023} for more details on the complexity associated to this technique). Each column contains the complexity associated to each of the three phases in the channel estimation process. Based on the considerations in \sectionname~\ref{sec:imperfect_knowledge}, the second-order statistics of the channel remain unchanged for $\tau_s$ channel coherence blocks, which means that the first two operations (i.e., estimation of $\widehat{\vect{Q}}^\mathsf{toe}$ and $\vect{A}_k$) need to be performed only \emph{once per $\tau_s$ channel realizations}, while the third operation (estimation of $\vect{A}_k\vect{y}_k$) is needed for each channel realizations. As a consequence, the whole estimation process has a complexity order given by the sum of the complexity due to $\widehat{\vect{Q}}^\mathsf{toe}$, plus the one due to $\vect{A}_k$, plus $\tau_s$\ times the one due to $\vect{A}_k\vect{y}_k$. Using \tablename~\ref{tab:complexity}, this means that the complexity order is $\mathcal{O}\left(MN^2+N^2N_\mathsf{V}+\tau_s N^2\right)$ for the \ac{MMSE}, and $\mathcal{O}\left(MN^2+N_\mathsf{H}^3+\tau_s \left(N_\mathsf{H}+N_\mathsf{V}\right)N\right)$ for the \ac{KBA}. Assuming a square array, in which $N_\mathsf{H}=N_\mathsf{V}=\sqrt{N}$, the complexity orders become $\mathcal{O}((M+\sqrt{N}+\tau_s )N^2)$ and $\mathcal{O}((MN+\sqrt{N}+2\tau_s\sqrt{N})N)$, respectively.

To better illustrate the impact of the \ac{UPA} size on the computational burden, \figurename~\ref{fig:complexity} reports the complexity as a function of the observation length $M$, using $N=256$ (dashed lines) and $N=4096$ solid lines), considering $\tau_s=25,000$. As can be seen, the gap between the \ac{MMSE} estimator (black lines) and the \ac{KBA} scheme (red lines) is already an order of magnitude for relatively small \acp{UPA} ($N=256$), while it becomes more significant for larger \acp{UPA} ($N=4096$), and keeps increasing with $N$. Note that, in highly time-varying scenarios, characterized by large values of $\tau_s$ (as the channel coherence time decreases faster than $T_s$), the gap increases. The same conclusions hold for other \ac{UPA} shapes other than the square one.

\figurename~\ref{fig:complexity} also reports the complexity required by a \ac{ULA} with the same number of antennas (note that, based on \tablename~\ref{tab:complexity}, the complexity does not depend on $M$). The computational saving is even more significant, at the cost of a less compact array size and a reduced capability to capture variations in both azimuth and elevation planes.

\begin{figure}[t]
  \begin{center}\vspace{-0.5cm}
    \subfigure[\acs{RZF} combining.]{
    {\includegraphics[width=\columnwidth]{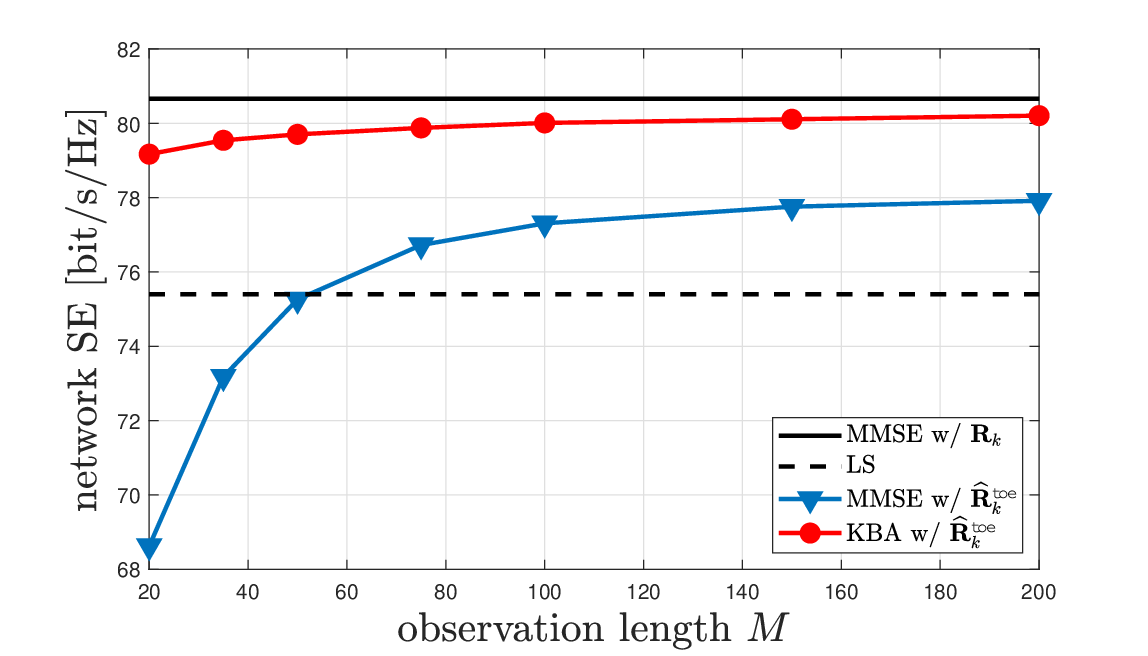}} 
    \label{fig10a}}
    \\
    \subfigure[\acs{MR} combining.]{
    {\includegraphics[width=\columnwidth]{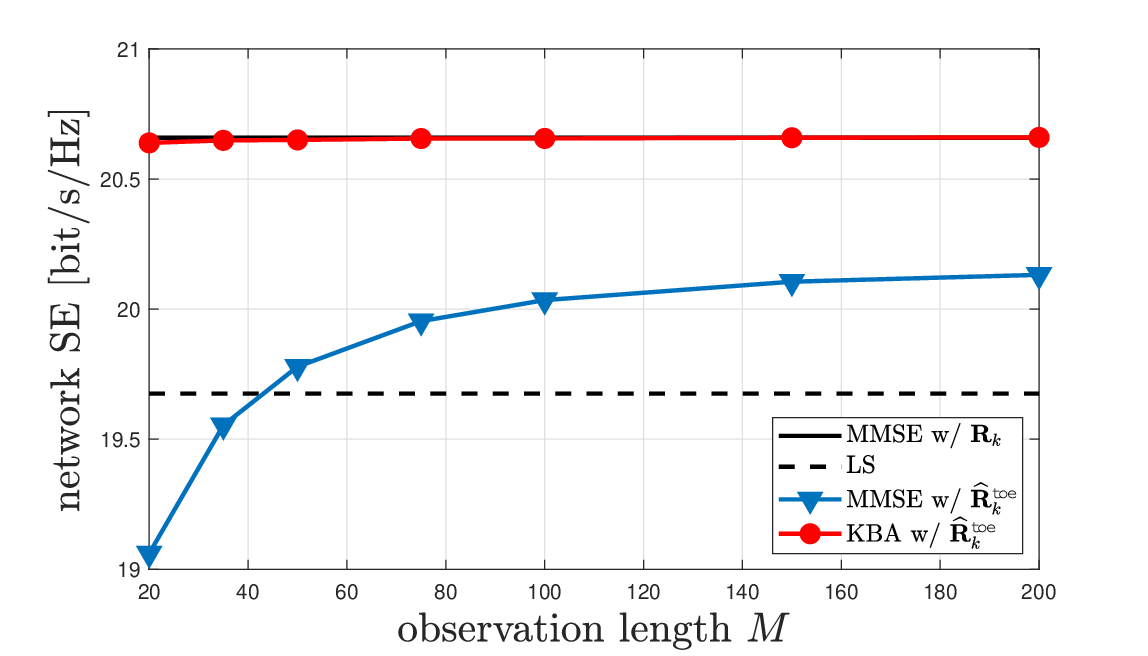}} 
    \label{fig10b}}
    \caption{Sum-\acs{SE} as a function of the observation length $M$ with \acs{RZF} and \acs{MF} combining.}\label{fig10}
  \end{center}
\end{figure}

\section{Uplink spectral efficiency evaluation}\label{sec:ul_sp}

Channel estimation is the ancillary task to support data detection, which is the ultimate goal of a communication system. 
Hence, we now compare the investigated solutions in terms of achievable uplink \ac{SE}. This is obtained with the well-known use-and-then-forget bound \cite[\sectionname~4.2]{massivemimobook}, which yields 
\begin{align}\label{eq:UaF-SE}
  \underline{\mathsf{SE}}_k = \left(1-\frac{\tau_p}{\tau_c}\right) \Ex{\log_2\left(1+\underline{\gamma}_k\right)}\qquad\textrm{[bit/s/Hz]},
\end{align}  
where the factor ${\tau_p}/{\tau_c}$ accounts for the fraction of samples per coherence block used for channel estimation, with $\tau_c$ being the length of each coherence block, and $\underline{\gamma}_{k}$ is given by
\begin{align}\label{eq:UaF-SINR}
  \frac{ \left| \Ex{ \vect{v}_k^\Htran \vect{h}_k } \right|^2 }
                         { \sum_{i=1}^{K}{ \Ex{\left| \vect{v}_k^\Htran \vect{h}_i \right|^2 } } 
                           - \left| \Ex{ \vect{v}_k^\Htran \vect{h}_k } \right|^2 
                           + \frac{\sigma^2}{\rho} \Ex{ \left\| \vect{v}_k \right\|^2 } },
\end{align}
with $\vect{v}_{k} \in \mathbb{C}^{N}$ being the receive combining vector associated to \ac{UE} $k$. 
Notice that the expectations are computed with respect to all sources of randomness. We consider both \ac{RZF} combining, according to which
\begin{align}\label{eq:MMSE_combiner}
  \vect{v}_k = \left( \sum_{i=1}^{K}{\widehat{\vect{h}}_i \widehat{\vect{h}}_i^\Htran} + \frac{\sigma^2}{\rho}  \vect{I}_{N} \right)^{-1} 
  \widehat{\vect{h}}_k
\end{align}
and \ac{MR} combining $\vect{v}_k=\widehat{\vect{h}}_k$. 



\figurename~\ref{fig10} shows the network \ac{SE}, given by 
\begin{align}\label{eq:UaF-sumSE}
  \underline{\mathsf{SE}} = \sum_{k=1}^K\underline{\mathsf{SE}}_k\qquad\textrm{[bit/s/Hz]}
\end{align}  
as a function of the observations $M$ used to estimate the channel statistics.
The \ac{RZF} combining scheme \eqref{eq:MMSE_combiner} is used in \figurename~\ref{fig10a}, while \ac{MR} is adopted in \figurename~\ref{fig10b}. The network parameters are detailed in \tablesname~\ref{tab:UPA} and \ref{tab:system}, with $K=10$ randomly placed \acp{UE}. The black solid line corresponds to the \ac{MMSE} estimator using perfect knowledge of the channel statistics $\vect{R}_k$ for both the combining and the estimation schemes (i.e., the estimation makes use of true $\vect{R}_k$), whereas the black dashed line depicts the results using the \ac{LS} estimator \eqref{eq:A-LS}; blue and red lines correspond to the results obtained assuming \emph{imperfect} channel statistics for the \ac{MMSE} and the \ac{KBA} methods, respectively, and thus $\vect{R}_k$ (used for both estimation and combining) is estimated using the methods introduced in \sectionname~\ref{sec:imperfect_knowledge}: more in detail, the \ac{MMSE} estimator uses \eqref{eq:Rsample} with $\eta=0.8$, whereas the \ac{KBA}-based estimator uses \eqref{eq:Rtoe} (\emph{without} any regularization parameter $\eta$). 


\begin{figure}[t]
  \begin{center}
    {\includegraphics[width=\columnwidth]{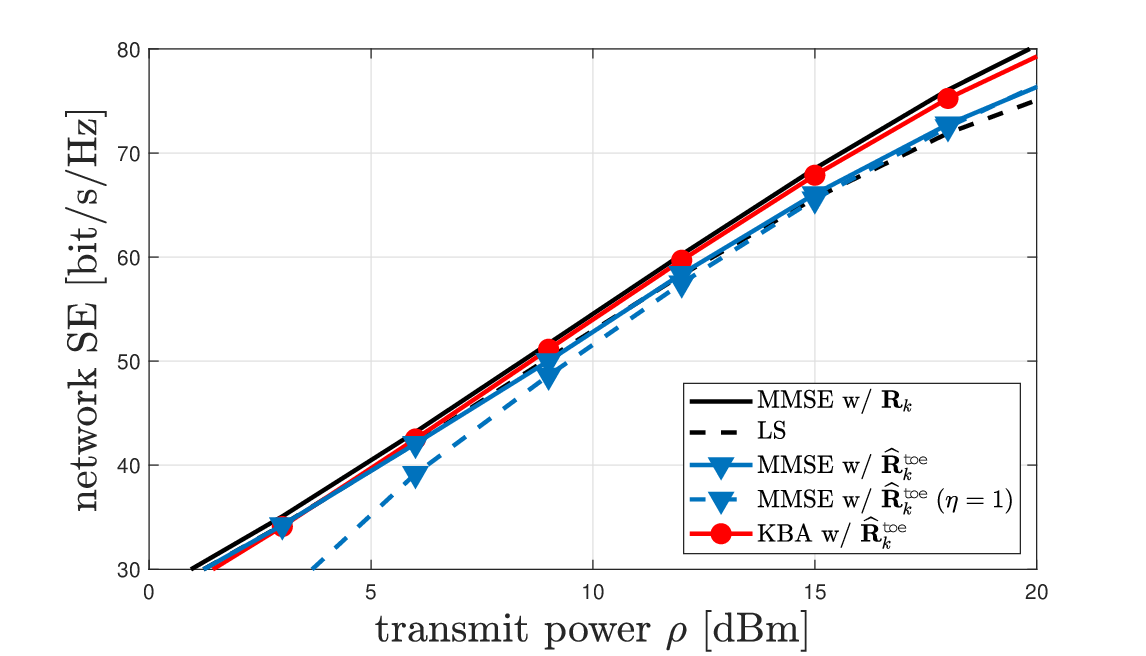}}
    \caption{Sum-\acs{SE} with \acs{RZF} as a function of the transmit power $\rho$ in dBm.}
    \label{fig11}
  \end{center}
\end{figure}


We notice that, while the gap between the imperfect-knowledge \ac{MMSE} estimator (blue line) and the benchmark curves remains significant even for large values of $M$, the performance of the \ac{KBA}-based scheme (red line) is very close to the optimal one (black curve), even for moderately low values of $M$ (for \ac{MR} combining, there is no difference, whereas the \ac{RZF} shows some gap, yet much smaller than that provided by the \ac{MMSE} estimator with imperfect knowledge, and decreasing with $M$). This confirms the results presented in \sectionname~\ref{imperfect_upa:performance}, in which the \ac{KBA}-based method, thanks to a simpler structure, introduces more robustness compared to the \ac{MMSE} counterpart for the same observation length $M$, and makes it more suitable for more dynamic scenarios, in which the coherence time is reduced (and hence $M$ needs to be kept as low as possible). In addition, the \ac{KBA}-based method does not call for any numerical optimization of the regularization parameter $\eta$ (unlike the \ac{MMSE} estimator -- see comments to \figurename~\ref{fig07}), and it can thus be universally adopted, irrespectively of the network setup (including \ac{UPA} parameters).


\figurename~\ref{fig11} shows the network \ac{SE} with \ac{RZF} as a function of the transmit power $\rho$. The network setup is identical to the one considered for \figurename~\ref{fig10a}, assuming $M=50$ channel observations for the covariance matrix estimation task. As can be seen, when $\rho=20\dBm$ (i.e., the same parameter setup considered for \figurename~\ref{fig10a}), the \ac{KBA}-based method outperforms the \ac{MMSE} one by $4.5\,\textrm{bit/s/Hz}$.\footnote{Note that the optimal regularization factor $\eta$ is a function of all simulation parameters, including $\rho$, and thus changes point by point -- this is the reason why the solid lines is always higher than the dashed line, without regularization.} Only when increasing $\rho$, in particular when $\rho\ge28\dBm$ (not reported here in the figure), in this simulation setup the two methods provide approximately the same performance. This means that the proposed \ac{KBA} method is particularly suitable in low-\ac{SNR} regimes, in which the link budget does not guarantee sufficient accuracy for the \ac{MMSE} counterpart. Similar trends, not reported here for the sake of brevity, are observed for \ac{MR} combining.


\section{Conclusion}\label{sec:conclusion}

We developed channel estimation schemes for large-scale \ac{MIMO} systems that achieve accuracy levels comparable to the optimal \ac{MMSE} estimator, while also offering reduced complexity and enhanced robustness to the imperfect knowledge of channel statistics. Both \acp{UPA} and \acp{ULA} were studied. In the case of \acp{UPA}, we exploited the inherent structure of the spatial correlation matrix to approximate it by a Kronecker product, and in the case of \acp{ULA}, we used a circulant approximation to further reduce the computational complexity. 
Comparisons were made with the optimal \ac{MMSE} estimator and other alternatives with less complexity (e.g., the \ac{LS} estimator). Numerical results confirmed the effectiveness of the proposed methods in terms of \ac{UL} \ac{SE} with various combining schemes, approaching \ac{MMSE} performance while significantly reducing computational complexity. Specifically, the computational load scales as $N\sqrt{N}$ and $N \log N$ for squared planar arrays and linear arrays, respectively. Furthermore, the proposed schemes exhibited improved robustness in scenarios with imperfect channel statistics and low-to-medium \ac{SNR} scenarios.
Future extensions of the proposed methodology may consider different shapes for the array (e.g., cylindrical), and the impact of near-field propagation conditions, which become relevant as the array size increases and/or the carrier frequency increases (e.g., see \cite{Cui2022, Huang24}).



\section*{Appendix}
The estimation of $\vect{h}_k$ calls for the product between $\vect{A}_{k}$ and $\vect{y}_{k}$. Accordingly, when $\vect{A}_{k}$ is written as a product of matrices (as occurs, for instance, with the \ac{MMSE} or the \ac{KBA} estimators), it is convenient not to perform the products. Indeed, assume that we have to compute $\vect{B}_{1} \vect{B}_{2} \vect{b}$, where $\vect{B}_{1}$ and  $\vect{B}_{2}$ are $N \times N$ matrices and  $\vect{b}$ is an $N$-dimensional vector. If we first multiply $\vect{B}_{1}$ and  $\vect{B}_{2}$, and then the resulting matrix and $\vect{b}$, we have to perform $N^3 + N^2$ multiplications and $(N-1)N^2+ (N-1)N$ additions. On the other hand, if we first compute the product between $\vect{B}_{2}$ and $\vect{b}$, and then the product between $\vect{B}_{1}$ and the resulting vector, $2N^2$ multiplications and $2(N-1)N$ additions are needed. This means, for example, that when considering the \ac{MMSE} estimator it is better to compute first $(\widehat{\vect{Q}}_{k}^{\mathsf{toe}})^{-1} \vect{y}_k$ and then $\widehat{\vect{R}}^{\mathsf{toe}}_k \left[(\widehat{\vect{Q}}_{k}^{\mathsf{toe}})^{-1} \vect{y}_k\right]$. An analogous approach can be conveniently used with the \ac{KBA} estimator. Another useful result, that can be exploited to reduce the complexity of the \ac{KBA} estimator, concerns the computation of the product $(\vect{C}_{1} \otimes \vect{C}_{2}) \vect{b}$, where $\vect{C}_{1}$ is $m \times m$, $\vect{C}_{2}$ is $n \times n$, and $\vect{b}$ is an $mn$-dimensional vector. The evaluation of $(\vect{C}_{1} \otimes \vect{C}_{2}) \vect{b}$ can efficiently be performed  with $(m+n)mn$ floating-point operations instead of $m^2n^2$ \cite{Dayar2015}.

\bibliographystyle{IEEEtran}
\bibliography{mybib}

\begin{IEEEbiography}
[{\includegraphics[width=1.0in,height=1.25in,clip,keepaspectratio]{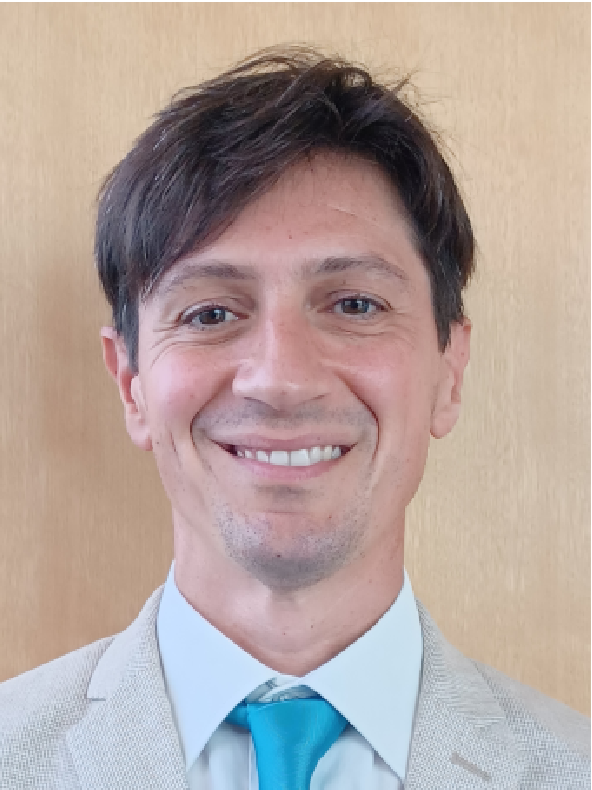}}]
{Giacomo Bacci} (Senior Member, IEEE) received the Ph.D. degree in information engineering from the University of Pisa, Pisa, Italy, in 2008. From 2006 to 2007, he was a visiting student research collaborator with Princeton University, Princeton, NJ, USA. From 2008 to 2014, he was a post-doctoral research fellow with the University of Pisa. From 2008 to 2012, he was also a software engineer with Wiser Srl, Livorno, Italy, and from 2012 to 2014, he was also enrolled as a visiting post-doctoral research associate with Princeton University. From 2015 to 2021, he was a product manager for interactive satellite broadband communications at MBI Srl, Pisa, Italy. Since 2022, he joined the University of Pisa as a tenure-track assistant professor.

Dr. Bacci is the recipient of the FP7 Marie Curie International Outgoing Fellowships for career development (IOF) 2011 GRAND-CRU, the Best Paper Award from the IEEE Wireless Communications and Networking Conference (WCNC) in 2013, the Best Student Paper Award from the International Waveform Diversity and Design Conference (WDD) in 2007, the Best Session Paper at the ESA Workshop on EGNOS Performance and Applications in 2005, and the 2014 URSI Young Scientist Award. He is currently serving as an associate editor for IEEE Communications Letters and EURASIP Journal on Advances in Signal Processing and he is a Senior Member of the IEEE and a Senior Member of the International Union of Radio Science.
\end{IEEEbiography}

\begin{IEEEbiography}
[{\includegraphics[width=1.0in,height=1.25in,clip,keepaspectratio]{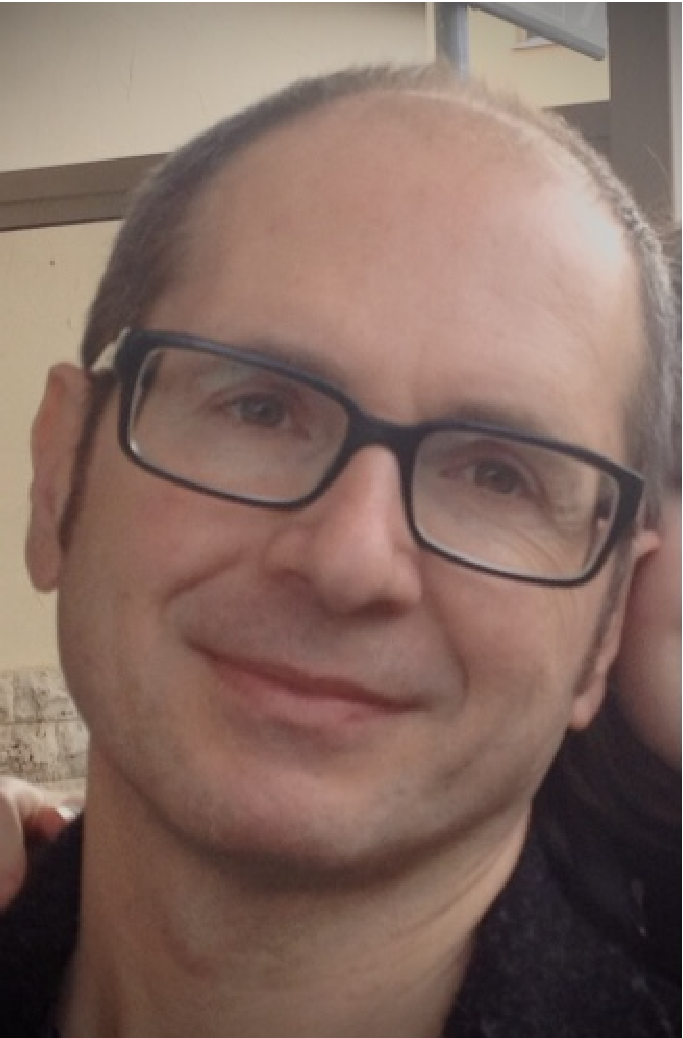}}]
{Antonio A. D'Amico} (Senior Member, IEEE) received the Laurea degree in Electronic Engineering in 1992 and the Ph.D. degree in 1997, both from the University of Pisa, Italy. He is currently an Associate Professor in the ``Dipartimento di Ingegneria dell'Informazione'' of the University of Pisa, Italy. His research interests are in digital communication theory and statistical signal processing, with emphasis on synchronization algorithms, channel estimation, localization, and detection algorithms.

He is currently serving as an Associate Editor for the IEEE Transations on Wireless Communications.
\end{IEEEbiography}

\begin{IEEEbiography}
[{\includegraphics[width=1.0in,height=1.25in,clip,keepaspectratio]{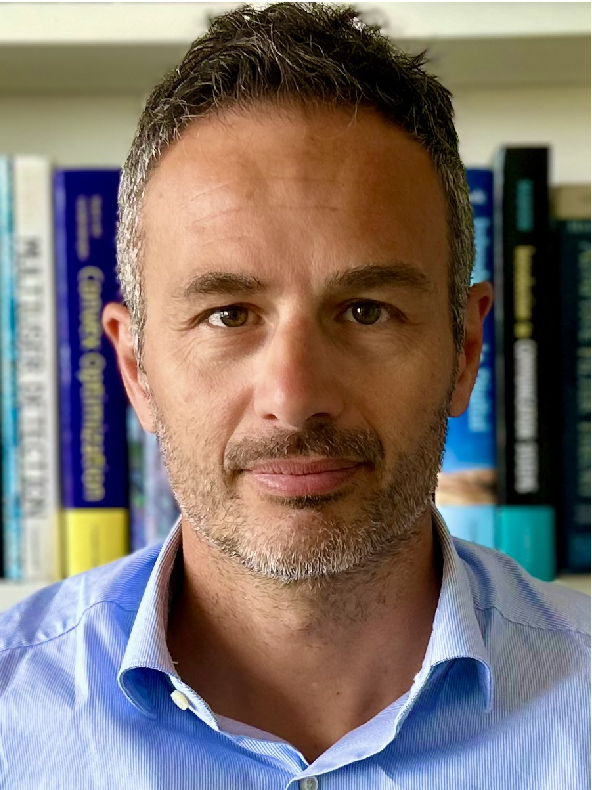}}]
{Luca Sanguinetti} (Senior Member, IEEE) received the Laurea degree  (cum laude) in Telecommunications Engineering and the Ph.D. degree in information engineering from the University of Pisa, Italy, in 2002 and 2005, respectively. In 2004, he was a visiting Ph.D. student at the German Aerospace Center (DLR), Oberpfaffenhofen, Germany. During the period June 2007 - June 2008, he was a postdoctoral associate in the Dept. Electrical Engineering at Princeton, NJ, USA. From July 2013 to October 2017 he was with Large Systems and Networks Group (LANEAS), CentraleSup\'elec, France. He is currently a Full Professor in the ``Dipartimento di Ingegneria dell'Informazione'' of the University of Pisa, Italy. 

He served as an Associate Editor for \textsc{IEEE Transactions on Communications}, \textsc{IEEE Transactions on Wireless communications} and \textsc{IEEE Signal Processing Letters}, and as Lead Guest Editor of \textsc{IEEE Journal on Selected Areas of Communications} Special Issue on ``Game Theory for Networks'' and as an Associate Editor for \textsc{IEEE Journal on Selected Areas of Communications} (series on Green Communications and Networking). Dr. Sanguinetti is currently a member of the Executive Editorial Committee of \textsc{IEEE Transactions on Wireless Communications}.

His expertise and general interests span the areas of communications and signal processing. Dr. Sanguinetti co-authored the textbooks \emph{Massive MIMO Networks: Spectral, Energy, and Hardware Efficiency} (2017) and \emph{Foundations on User-centric Cell-free Massive MIMO} (2020). He received the \emph{2018 and 2022 Marconi Prize Paper Award in Wireless Communications}, the \emph{2023 IEEE Communications Society Outstanding Paper Award}. 
\end{IEEEbiography}

\end{document}